\documentclass[preprint,12pt]{qrei}

\usepackage{amssymb}
\usepackage{graphicx}
\usepackage{amsmath}
\usepackage{amsthm}
\usepackage{marvosym}
\newtheorem{mydef}{Definition}

\usepackage{hyperref}
\usepackage{url}

\begin{document}

\begin{frontmatter}



\title{Formal Specification and Quantitative Analysis of a Constellation of Navigation Satellites}

 \author[label1,label3]{Zhaoguang Peng}
  \address[label1]{China Ceprei Laboratory, Guangzhou, China}
  \address[label2]{School of Computing Science, University of Glasgow, Glasgow, UK}
 \address[label3]{School of Reliability and Systems Engineering, Beijing University of Aeronautics and Astronautics, Beijing, China}
  \author[label2]{Yu Lu\corref{cor1}}
  \cortext[cor1]{Correspondence to: Yu Lu, School of Computing Science, University of Glasgow, Glasgow G12 8RZ, UK. Email: y.lu.3@research.gla.ac.uk}
  \author[label2]{Alice Miller}
  \author[label3]{Tingdi Zhao}
  \author[label2]{Chris Johnson}



\begin{abstract}
Navigation satellites are a core component of navigation satellite based systems such as GPS, GLONASS and Galileo which provide location and timing information for a variety of uses. Such satellites are designed for operating on orbit to perform tasks and have lifetimes of 10 years or more. Reliability, availability and maintainability (RAM) analysis of systems has been indispensable in the design phase of satellites in order to achieve minimum failures or to increase mean time between failures ($MTBF$) and thus to plan maintenance strategies, optimise reliability and maximise availability. In this paper, we present formal models of both a single satellite and a navigation satellite constellation and logical specification of their reliability, availability and maintainability properties respectively. The probabilistic model checker $\mathsf{PRISM}$ has been used to perform automated analysis of these quantitative properties.
\end{abstract}


\end{frontmatter}


\section{Introduction}

With the emergence of efficient, high-performance, and low cost satellites, earth orbiting satellites are often deployed in satellite constellations and space systems to ensure reliable and dependable missions. These kinds of satellites have played an essential part in both civil and military contexts, and support a wide range of applications ranging from satellite navigation to space stations. Most of these applications are not only safety-critical but also mission-critical, thus they heavily depend on such infrastructures within the systems. A group of artificial satellites which work in concert is known as a satellite constellation. A satellite constellation is a number of satellites with coordinated ground coverage, operating together under shared control, synchronised so that they overlap in coverage and complement rather than interfere with other satellites coverage.

A navigation satellite system is a satellite constellation consisting of a number of navigation satellites that provide autonomous geospatial positioning with global or regional coverage. It is by far one of the most successful applications of satellites, and has been developed since 1973. A navigation satellite system with global coverage is referred to as a global navigation satellite system (GNSS). Leading international projects include the United States' Global Position System (GPS) and Russia's Global Navigation Satellite System (GLONASS), both of which are fully operational GNSSs. In addition, China is expanding its regional Beidou navigation system into the global compass navigation system, and the European Union's Galileo positioning system is a GNSS in the initial deployment phase. Both of these systems are planned to be fully operational in the next decade. Other countries such as India, France, and Japan are in the process of developing their own regional navigation systems. See \cite{WLW07} for a good overview of navigation satellite systems.

A satellite is designed to a functional requirement and it is important that it satisfies this requirement. However it is also desirable that the satellite should be predictably available and this depends upon its reliability and availability. We aim to help the military or civil end users of the satellite to assess the likelihood and consequences of fault or failure to their operations. Reliability, availability and maintainability (RAM) analysis has been indispensable in the design phase of navigation satellite systems in order to achieve minimum failures or to increase mean time between failures ($MTBF$) and thus to plan maintenance strategies, optimise reliability and maximise availability. The question of how to select optimal configurations and maintenance plans and underlying resources, to satisfy requirements and improve efficiency is a key research question. This concern calls for effective solutions to the challenges of verifying large and complex navigation systems.

Until now, attempts to verifying satellite based systems have been piecemeal. Verification largely depends on brute force approaches, such as simulation and testing. Simulation is the common testing and validation approach used for the verification of satellite systems. Given a system, a finite subset of the possible scenarios are selected in a specific simulation environment, and then statistical analysis techniques are applied to obtain probabilistic results for that system. However, simulation has been unable to keep up with the growth in design complexity of satellite systems. On the other hand, formal verification is a well-established technique in Computer Science for either detecting errors, or for providing increased confidence in the reliability of a system. It is therefore timely to apply formal verification techniques to this domain. Formal verification can be applied to formally verify satellite systems using automated tools including model checkers or theorem provers.

Model checking is a formal verification technique that involves defining a model of a system from a formal specification. The model is then used to check desired properties of the system. This involves exploring the underlying state space of the model, and specifying properties via some formal logic such as temporal logic. In this context, the effects of proposed changes to an on orbit system can be first checked via a model, rather than via expensive prototypes. The required reliability, availability, and maintainability properties of satellite systems can be expressed in temporal logic, and so lend themselves very well to proof via model checking.

The goal of the paper is to adopt probabilistic model checking to cope with the verification demand introduced by satellite systems. Probabilistic model checking is a formal method for specifying quantitative properties of a system model. Models obtained by this technique are normally extensions or variants of Markov chains or automata, extended with costs and rewards that estimate resources and their usage during operation. Properties to be verified or analysed are specified in temporal logic with auxiliary operators such as probability and reward. We present an automated quantitative analysis of reliability, availability, and maintainability of both a single satellite system and a navigation satellite system, using the probabilistic model checker $\mathsf{PRISM}$ \cite{KNP09}.

Our paper is organised as follows. In Section \ref{sect:system} we describe the underlying navigation satellite systems. In Section \ref{sect:formalmethods} the use of formal methods is introduced, while in Section \ref{sect:pmc} we give technical background on probabilistic model checking. In Section \ref{sect:fmc} we present our formal specifications of a single satellite and constellation systems and their associated continuous-time Markov chain models respectively. Then, we analyse reliability, availability, and maintainability using the probabilistic model checker $\mathsf{PRISM}$ for a single satellite and a satellite constellation in Sections \ref{sect:lspar}. In Section \ref{sect:relatedWork} we report related work for verifying satellite systems using model checking. Finally, in Section \ref{sect:conclusion} we conclude and outline directions for future research.


\section{Satellite systems}\label{sect:system}

As an important application of satellite constellation, navigation satellite systems consist of three major segments: a space segment, a control segment, and a user segment. The space segment is made up of a number of satellites, and is responsible for sending the navigation signal on the specific frequency. It is constantly orbiting the surface at an altitude of approximate three earth radii, and emitting signals that travel at approximately the speed of light. The control segment monitors the health and status of the space segment and controls the state of satellites, and updates the data of those satellites. The user segment consists of antennas and receiver processors, which receive the signals broadcasted by the satellites and decode them to provide precise information about the receiver’s position and velocity.

In a satellite constellation, fault or failure of more than one satellite will have a direct impact on the stable state of the space geometry and temporal relationship, and the performance of the constellation. So the performance of the constellation is a direct consequence of the state of the constellation. Therefore, the state of the constellation has a close relationship with the state of every satellite in the constellation. So each satellite is critical to the constellation.

In this paper, our task is to help the end users of satellite based systems to evaluate the probability and consequences of faults or failures. The terms of fault and failure in our context can be defined according to \cite{Czi13} as follows:

\begin{itemize}
\item Fault: the condition of a satellite that occurs when one of its components or assemblies degrades or exhibits abnormal behaviour;
\item Failure: the termination of the ability of a satellite to perform a required function.
\end{itemize}

Failure is an event as distinguished from fault, which is a state. According to \cite{Czi13}, the failure mode is the result by which a failure is observed. After a failure, a satellite in the constellation will be systematically examined in order to identify the failure mode, and to determine the nature of the failure and its basic cause. There are three kinds of failure mode of the satellite: long-term failure (unrecoverable failure), short-term failure, and Operations and Maintenance (O\&M) failure. These failure modes are described as follows:

\begin{itemize} 
\item Long-term failure: this failure is vital to the satellite. If a long-time failure has happened, it usually needs to launch another satellite to replace the failed one. Practically, it indicates that the failed satellite is at the end of its life. It has also been called wear out failure;
\item Short-term failure: this refers to a failure that can be repaired in several hours or days. This kind of failure mode means that there is usually no need to launch a new satellite to replace the failed satellite;
\item O\&M failure: is due to planned maintenance operations, such as navigation satellite orbit manoeuvre and atomic clock switching. We usually do not consider the outage time that is induced by these operations as a failure. It is not expected to impact the continuity of the constellation, but the performance of the constellation.
\end{itemize}

\begin{figure*}[ht]
\begin{center}
\includegraphics[width=14cm]{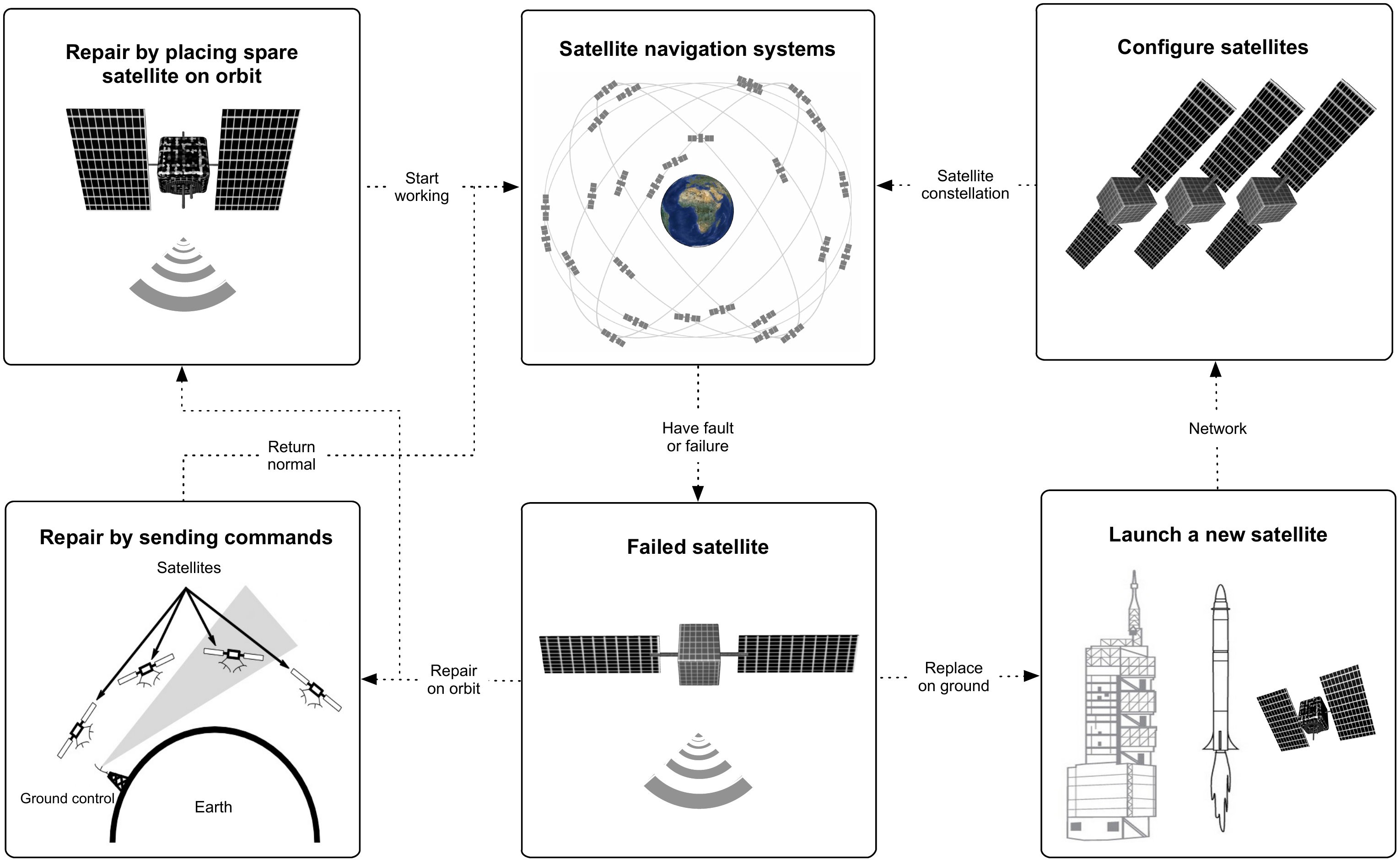}
\caption{An overview of navigation satellite systems.}
\label{fig:system}
\end{center}
\end{figure*}

Whenever a satellite has a fault or fails, there is a chance to repair the satellite on orbit by, for example, rebooting the satellite system, updating the satellite software, or switching the orbit of the satellite. There are three satellite backup modes available for maintenance strategies: on orbit backup, parking orbit backup, and Launch on Need (LON). The on orbit backup mode and parking orbit backup mode are further referred to as space backup. In this paper, we consider both space backup and LON backup. The main navigation satellite system to be modelled and analysed is depicted in Figure~\ref{fig:system}.

Satellites deployed at the parking orbit backup mode can also be used to work with on orbit satellites. For LON backup mode, it usually takes several months to replace failed satellite, while for space backup mode it only takes one or two days. Because of the lower mean time to repair ($MTTR$) for the space backup mode, it has been widely applied in most constellation projects. In the GPS project, the redundant satellites are working with on orbit satellites, so failed satellites can be replaced in a short time.


\section{Analysis techniques}\label{sect:formalmethods}

In this paper, our models are Continuous-time Markov Chains (CTMCs), and we verify our models using model checking. Before formally introducing this technique and discussing the role of formal verification, we briefly review some traditional software and hardware verification and analysis techniques that can be applied to analysing satellite based systems, which are led by testing and simulation.

Testing is a dynamic verification technique that involves actually running software systems. Testing takes the system under analysis and uses inputs as tests. Correctness is thus verified by running the system to traverse a set of execution paths. Based on the results during test execution, the actual output of the system is compared to the system specification which is usually in the form of documents.

Simulation is similar to testing, but is applied to system models which represent the underlying system for analysis. Models are usually described using hardware description languages. A simulator is used to examine execution paths of the system model based on configuration inputs. These inputs can be provided by a user, or by automated approaches such as using a random generator. A mismatch between the simulator's result and the specification of the system exhibits the incorrect behaviours.

Both of these verification techniques are limited in that they only allow exploration of a small subset of many possible scenarios. Formal methods is the application of mathematical modelling and reasoning to prove that an implementation coincides with precisely expressed notion of formal specification. In this context, the purpose of formal analysis and verification is to analyse the performance and to verify the correctness and properties of satellite based systems in such a way that faults and failures can be identified. Model checking and theorem proving are formal techniques that can be used to detect faults and failures in a formal specification.

Although historically these forms of verification were used to prove correctness of explicit software and hardware designs, these days they are also used for failure analysis. They are generally applied during the design phase, where they are arguably most effective, for verifying correctness and other essential properties. Model checking is an automated analysis technique that requires expert knowledge to use. The user must provide an initial specification of the system itself, as well as logical properties describing its desired behaviour.

One strength of model checking to traditional analysis techniques is that it is not sensitive to the probability that a fault or failure is exposed; this contrasts with testing and simulation that are aimed at tracing the most probable faults or failures. Moreover, it allows one to precisely analyse results of checking desired properties. Model checking is a general analysis technique that is applicable to a wide range of applications such as embedded systems, software engineering, and hardware design. It also supports analysing properties individually, thus allowing one to focus essential properties first. This enables incomplete formal models to be specified and verified.

The formal model of systems can be defined using a high-level formalism or extracted directly form software using methods such as abstract interpretation. Verification involves checking paths of the state transition graph (or state-space) of the model. Traditionally this involves either exhaustive or on-the-fly search of the state-space in which states are stored explicitly. Another method – - symbolic model checking \cite{CGP99} - – involves search of a symbolic representation of the state space, in which groups of states and transitions are explored in a single step.

Quantitative verification is a analysis technique for establishing quantitative properties of a system model. Models analysed through this method are typically variants of Markov chains, annotated with costs and rewards that describe resources and their usage during execution. Properties are expressed in temporal logic extended with probabilistic and reward operators. Quantitative verification involves a combination of a traversal of the state transition system of the model and numerical computation. In this paper, we employ the power of probabilistic model checking, which is a leading quantitative verification and analysis technique for a wide variety of systems.


\section{Probabilistic model checking}\label{sect:pmc}

In this section we introduce some formal notation that is relevant to probabilistic model checking. Our definitions in Section 4.1 and 4.2 follow \cite{BK08}, from which further details can be found.

\subsection{Continuous-time Markov chains}

Our approach is event based because of the fault and failure events that can be sensed and monitored in the satellite systems. Rates are assigned to events and our underlying semantics is continuous time Markov chains (CTMCs): the state space is discrete but time is continuous. In this section, we briefly review the basic concept of CTMCs.

\begin{mydef}\label{def:ctmc}
Let $AP$ be a fixed, finite set of atomic propositions. Formally, a continuous-time Markov chain (CTMC) $\mathcal{C}$ is a tuple ($S$,$s_{init}$,$R$,$L$) where:
\begin{itemize}
\item $S=\{s_{1},s_{2},...,s_{n}\}$ is a finite set of states.
\item $s_{init}\in S$ is the initial state.
\item $R:\ S\times S\rightarrow \mathbb{R}_{\geq 0}$ is the transition rate matrix.
\item $L:\ S\rightarrow 2^{AP}$ is a labelling function which assigns to each state $s_{i}\in S$ the set $L(s_{i})$ of atomic propositions $a\in AP$ that are valid in $s_{i}$.
\end{itemize}
\end{mydef}

Intuitively, $R(s_{i},s_{j})>0$ if and only if there is a transition from state $s_{i}$ to state $s_{j}$. Furthermore, $R(s_{i},s_{j})$ specifies that the probability of moving from $s_{i}$ to $s_{j}$ within $t$ time units is $1-e^{-R(s_{i},s_{j})\cdot t}$, an exponential distribution with rate $R(s_{i},s_{j})$. If $R(s_{i},s_{j})>0$ for more than one state $s_{j}$, a competition between the transitions originating in $s_{i}$ exists, known as the race condition. 

The probability to move from a non-absorbing state $s_{i}$ to a particular state $s_{j}$ within $t$ time units, i.e., the transition $s_{i}\rightarrow s_{j}$ wins the race, is given by:

\begin{equation}
P(s_{i},s_{j},t)=\frac{R(s_{i},s_{j})}{E(s_{i})}\cdot(1-e^{-E(S_{i})\cdot t}),
\end{equation}

\noindent where $E(s_{i}) =\sum_{s_{j}\in S}R(s_{i}, s_{j})$ denotes the total rate at which any transition outgoing from state $s_{i}$ is taken. More precisely, $E(s_{i})$ specifies that the probability of taking a transition outgoing from the state $s_{i}$ within $t$ time units is $1-e^{-E(S_{i})\cdot t}$, since the minimum of two exponentially distributed random variables is an exponentially distributed random variable with rate the sum of their rates. Consequently, the probability of moving from a non-absorbing state $s_{i}$ to $s_{j}$ by a single transition, denoted $P(s_{i},s_{j})$, is determined by the probability that the delay of going from $s_{i}$ to $s_{j}$ finishes before the delays of other outgoing edges from $s_{i}$; formally, $P(s_{i},s_{j}) = R(s_{i},s_{j})/E(s)$. For an absorbing state $s_{i}$, the total rate is $E(s_{i})$. In that case, we have $P(s_{i},s_{j})=0$ for any state $s_{j}$.

\subsection{Continuous stochastic logic}

The probabilistic model checker $\mathsf{PRISM}$ provides support for automated analysis of a wide range of quantitative properties, such as \textquotedblleft what is the probability of a failure causing the satellite to stop working within 12 hours?\textquotedblright, \textquotedblleft what is the worst-case probability of the satellite on-board system terminating due to an error, over all possible initial configurations?\textquotedblright, or \textquotedblleft what is the worst-case expected time taken for the satellite signal to be received?\textquotedblright.

In this paper, we use Continuous Stochastic Logic (CSL) \cite{ASS+96,BKH99} to specify reliability, availability, and maintainability properties. CSL is inspired by the logic Computation Tree Logic (CTL) \cite{Eme90}, and its extensions to discrete time stochastic systems (PCTL) \cite{HJ94}, and continuous time non-stochastic systems (TCTL) \cite{ACD90}. There are two types of formulae in CSL: state formulae, which are true or false in a specific state, and path formulae, which are true or false along a specific path.

\begin{mydef}\label{def:csl}
Let $a\in AP$ be an atomic proposition, $p\in[0,1]$ be a real number, $\bowtie\ \in\{\leq,<,>,\geq\}$ be a comparison operator, and $I\subseteq\mathbb{R}_{\geq 0}$ be a non-empty interval. The syntax of CSL formulas over the set of atomic propositions $AP$ is defined inductively as follows:
\begin{itemize}
\item $true$ is a state-formula.
\item Each $a\in AP$ is a state formula.
\item If $\Phi$ and $\Psi$ are state formulas, then so are $\neg\Phi$ and $\Phi\wedge\Psi$.
\item If $\Phi$ is state formula, then so is $\mathcal{S}_{\bowtie p}(\Phi)$.
\item If $\varphi$ is a path formula, then $\mathcal{P}_{\bowtie p}(\varphi)$.
\item If $\Phi$ and $\Psi$ are state formulas, then $\mathcal{X}_{I}\Phi$ and $\Phi\mathcal{U}_{I}\Psi$ are path formulas.
\end{itemize}
\end{mydef}

Formula $\mathcal{S}_{\bowtie p}(\Phi)$ asserts that the steady-state probability for a state satisfying $\Phi$ meets the bound $\bowtie p$. Similarly, formula $\mathcal{P}_{\bowtie p}(\varphi)$ asserts that the probability measure of the paths satisfying $\varphi$ meets the bound given by $\bowtie p$. The operator $\mathcal{P}_{\bowtie p}(.)$ replaces the usual CTL path quantifiers $\exists$ and $\forall$. Intuitively, $\exists\varphi$ represents that there exists a path for which $\varphi$ holds and corresponds to $\mathcal{P}_{>0}(\varphi)$, and $\forall\varphi$ represents that for all paths $\varphi$ holds and corresponds to $\mathcal{P}_{>1}(\varphi)$. The temporal operator $\mathcal{X}_{I}$ is the timed variant of the standard next operator in CTL; the path formula $\mathcal{X}_{I}\Phi$ asserts that a transition is made to a $\Phi$ state at some time point $t\in I$. Operator $\mathcal{U}_{I}$ is the timed variant of the until operator of CTL; the path formula $\Phi\mathcal{U}_{I}\Psi$ asserts that $\Psi$ is satisfied at some time instant in the interval $I$ and that at all preceding time instants $\Phi$ holds.

One of the most important operators is the $P$ operator, which is used to reason about the probability of an event. This operator was originally proposed for use in the logic PCTL but also features in the other logics supported by $\mathsf{PRISM}$, such as CSL. The $P$ operator is applicable to all types of models supported by $\mathsf{PRISM}$.

It is often useful to compute the actual probability that some behaviour of a model is observed. Therefore, $\mathsf{PRISM}$ allows a variation of the $P$ operator to be used in a query, i.e., $P_{=?}[pathprop]$, which returns a numerical rather than a Boolean value (i.e., the probability that $pathprop$ is true). In our paper, we are interested in directly specifying reliability, availability, and maintainability properties which evaluate to a numerical value. For example, we might wish to calculate the probability that process 1 terminates before process 2 does (say). This can be specified as $P_{=?} [ !proc2\_terminate\ U\ proc1\_terminate ]$, where $U$ is the \textquotedblleft until\textquotedblright\ temporal operator.

Another important operator we use is the $R$ operator, which specifies a cumulative reward property that associate a reward with each path of a model, but only up to a given time bound. The property $R_{=?}[ C<=t ]$  corresponds to the reward cumulated along a path until $t$ time units have elapsed. For CTMCs, the bound $t$ can evaluate to a real value. Some typical examples of properties using $P$ and $R$ operators can be found on the Property Specification section of the $\mathsf{PRISM}$ website.

\subsection{Reactive modules of $\mathsf{PRISM}$}

$\mathsf{PRISM}$ supports the analysis of several types of probabilistic models: discrete-time Markov chains (DTMCs), continuous-time Markov chains (CTMCs), Markov decision processes (MDPs), probabilistic automata (PAs), and also probabilistic timed automata (PTAs), with optional extensions of costs and rewards \cite{KNP09}. Moreover, $\mathsf{PRISM}$ allows us to verify properties specified in the temporal logics PCTL for DTMCs and MDPs and CSL for CTMCs. Models are described using the $\mathsf{PRISM}$ language, a simple, state-based language.

Markov models to be verified using specified in $\mathsf{PRISM}$ are specified using the $\mathsf{PRISM}$ modelling language which is based on the Reactive Modules formalism \cite{AH99}. A fundamental component of this language is a {\it module}. A system is constructed as the parallel composition of a number of modules. A module is specified as:
\[
\textbf{module}\ name\ ...\ \textbf{endmodule}
\]

A module definition consists of two parts: one containing variable declarations, and the other {\it commands}. At any time, the {\it state} of a model is determined by the current value of all of the variables of all of the components (modules). A variable declaration has the form:
\[
x\ :\ [0..2]\ \textbf{init}\ 0;
\]

In this example, variable $x$ is declared, with range $[0..2]$ and initial value $0$. The behaviour of each module is specified using commands, comprising a guard and one or more updates of the form:
\[
[action]\ guard\ \rightarrow\ rate\ :\ update
\]
or,
\[
[action]\ guard\ \rightarrow\ rate_1:update_1+rate_2:update_2+...
\]

The (action) label is optional, and is used to force two or more modules to synchronise. Updates in commands are labelled with positive-valued rates \cite{KNP09} for CTMCs. The $+$ indicates the usual non-deterministic choice. Within a module, multiple transitions can be specified either as several different updates in a command, or as multiple commands with overlapping guards. The following examples:
\[
\begin{array}{l}
   \ [\ ]\ x=0\ \rightarrow\ 0.5:(x'=0);\\
   \ [\ ]\ x=0\ \rightarrow\ 0.8:(x'=1);
\end{array}
\]
and
\[
[\ ]\ x=0 \rightarrow\ 0.5:(x'=0)\ +\ 0.8:(x'=1);
\]
are equivalent. The guard $x=0$ indicates that command is only executed when variable $x$ has value 0. The updates $(x'=0)$ and $(x'=1)$ and their associated rates indicate that the value of $x$ will remain at 0 with rate 0.5 and change to 1 with rate 0.8. In a CTMC, when multiple possible transitions are available in a state, a race condition occurs \cite{KNP07}. The rate of the synchronised transition is the product of all the individual rates.


\section{Formal specification of satellite systems}\label{sect:fmc}

In this section, we give an description of the basic formal models of both a single satellite and a constellation of navigation satellites.

\subsection{A formal model of a single satellite}

\begin{figure*}[ht]
\begin{center}
\includegraphics[width=14cm]{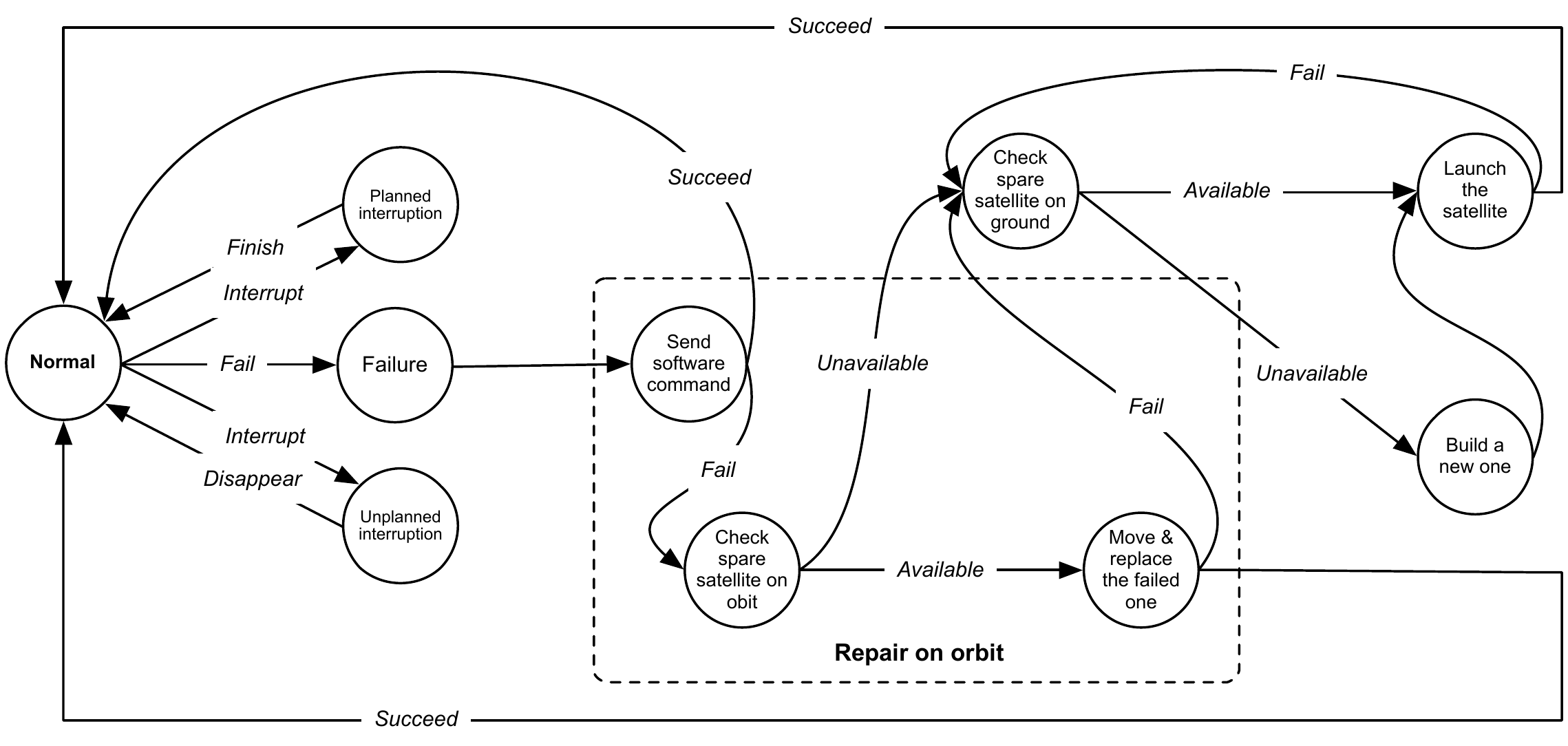}
\caption{A reference model of a single satellite.}
\label{fig:model}
\end{center}
\end{figure*}

The abstract model of a single satellite is illustrated in Figure~\ref{fig:model}, parameters are omitted. We take a CTMC as our underlying $\mathsf{PRISM}$ model for our abstract model.

We specify our CTMC model with states, a transition rate matrix, and a labelling function. Initially, the satellite runs in the normal state. After a period of execution it could be interrupted by an planned or unplanned interruption. Planned interruptions are normally caused by certain types of Operations and Maintenance (O\&M), which could include manoeuvring the station, atomic clock maintenance, software updates, and hardware maintenance. Unplanned interruptions can be caused by solar radiation, the earth's magnetic field cosmic rays, which result in a satellite Single Event Upset (SEU). However, both planned and unplanned interruptions are usually temporary, lasting just several hours. An unplanned interruption usually disappears automatically. The satellite can fail at any time during its lifetime due to End-of-Life (EOL) outage or other vital failures.

When the satellite fails, staff on the ground must decide upon the best approach to repair it. It may be possible that failures can be resolved on orbit by giving specific software commands to the satellite. Otherwise it might be necessary to move a redundant satellite into position to replace the failed satellite. If no redundant satellite is available then a new satellite must be manufactured and launched. In the worst case the new satellite does not launch successfully due to a known probability of satellite launch failure.

Most of our parameter values correspond to those of the latest United States' GPS system, GPS Block III satellites. The GPS III series is the newest block of GPS satellites. GPS III provides more powerful signals than previous versions in addition to enhanced signal reliability, accuracy, and integrity. The key improvement is the 15 years' design lifespan \cite{NOA13}. Due to privacy and secrecy reasons, NASA does not release all actual data of GPS III that we need in our analysis. Thus, in order to perform the analysis convincingly, we use some generic data of some very similar satellites instead. We believe this this will not result in a loss of generality since all data come from real satellites.

\begin{figure}[ht]
\begin{center}
\setlength{\fboxrule}{0.5pt} 
\setlength{\fboxsep}{0.25cm} 
\fbox{\includegraphics[width=14cm]{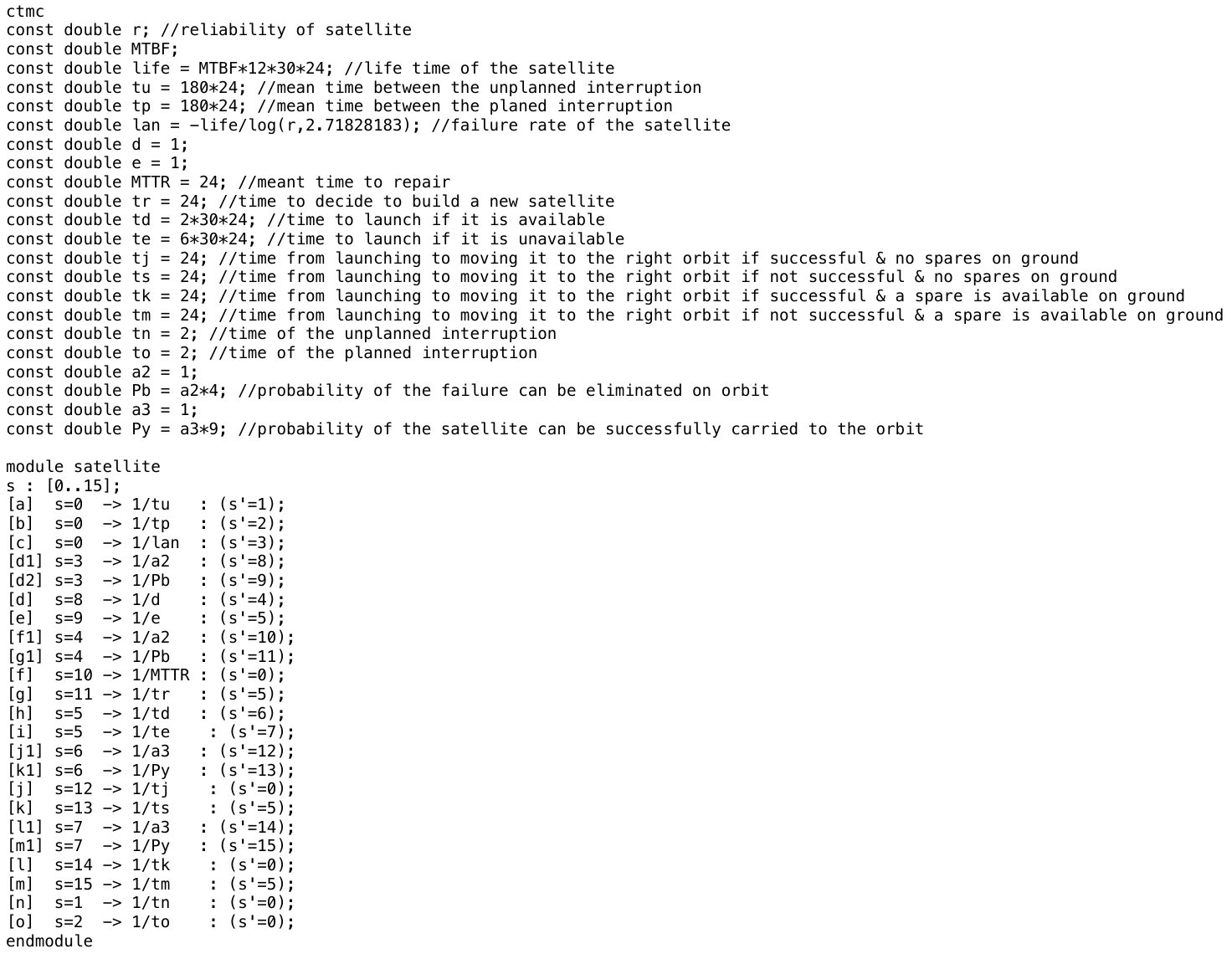}}
\caption{$\mathsf{PRISM}$ module for a single satellite.}
\label{fig:code1}
\end{center}
\end{figure}

All parameters used in our CTMC model and properties are specified in Table~\ref{tab:one}, and are described as follows. We use $p$ to express probability and $t$ for time, and the reliability of the satellite is $r$. If the satellite fails, we say that it moves from a \textquotedblleft normal\textquotedblright\ state to a \textquotedblleft failure\textquotedblright\ state. The mean time to unplanned interruption is $t_{u}$, while the mean time to planned interruption is $t_{p}$. When the satellite fails, the probability of the failure being resolved on orbit by moving a redundant satellite to replace the failed one is $p_{b}$. If on orbit repair is not possible, a new satellite is needed. The time taken to decide to build a new satellite and for one to be manufactured are $t_{r}$ and $t_{d}$ respectively. If a new satellite is to be manufactured, the probability of successful launch is $p_{y}$. After successful launch, the time taken for the satellite to move to the right position and a normal signal sent from it to be received on the ground is $t_{k}$. Our $\mathsf{PRISM}$ specification is given in Figure~\ref{fig:code1}.

Specifically, in Figure~\ref{fig:code1} $tj$ is the time from launching the satellite to moving it to the right orbit when the satellite has been successfully carried to the orbit if there are no spare satellites on the ground, and $ts$ is the time from launching the satellite to moving it to the right orbit when the satellite has not been carried to the right orbit if there are no spare satellites on the ground, and $tk$ the time from launching the satellite to moving it to the right orbit when the satellite has been successfully carried to the orbit if spare satellites is available on the ground, and $tm$ is the time from launching the satellite to moving it to the right orbit when the satellite has not been carried to the right orbit if spare satellites is available on the ground.

\subsection{A formal model of a constellation of navigation satellites}

We have modelled a single satellite as a CTMC, by specifying it in $\mathsf{PRISM}$. However, the RAM analysis of a single satellite appears insufficient for larger navigation satellite systems. For a large global navigation system, at least 24 satellites are required. Even for a regional navigation system, at least 4 satellites are required. Our $\mathsf{PRISM}$ model for a satellite constellation is thus constructed using our specification for a single satellite, with a number of modifications as follows:

\begin{itemize}
\item the number of satellites is declared as a global variable, and multiple satellite modules are instantiated;
\item the configuration of the satellite constellation is defined;
\item redundant satellites that are usually called spare satellites are included.
\end{itemize}

Note that the last modification above is due to the fact that, in a real system, if an on orbit satellite fails, redundant on orbit satellites are used to move and replace them, to ensure the availability of the constellation.

\begin{figure*}[ht]
\centerline{\includegraphics[width=14cm]{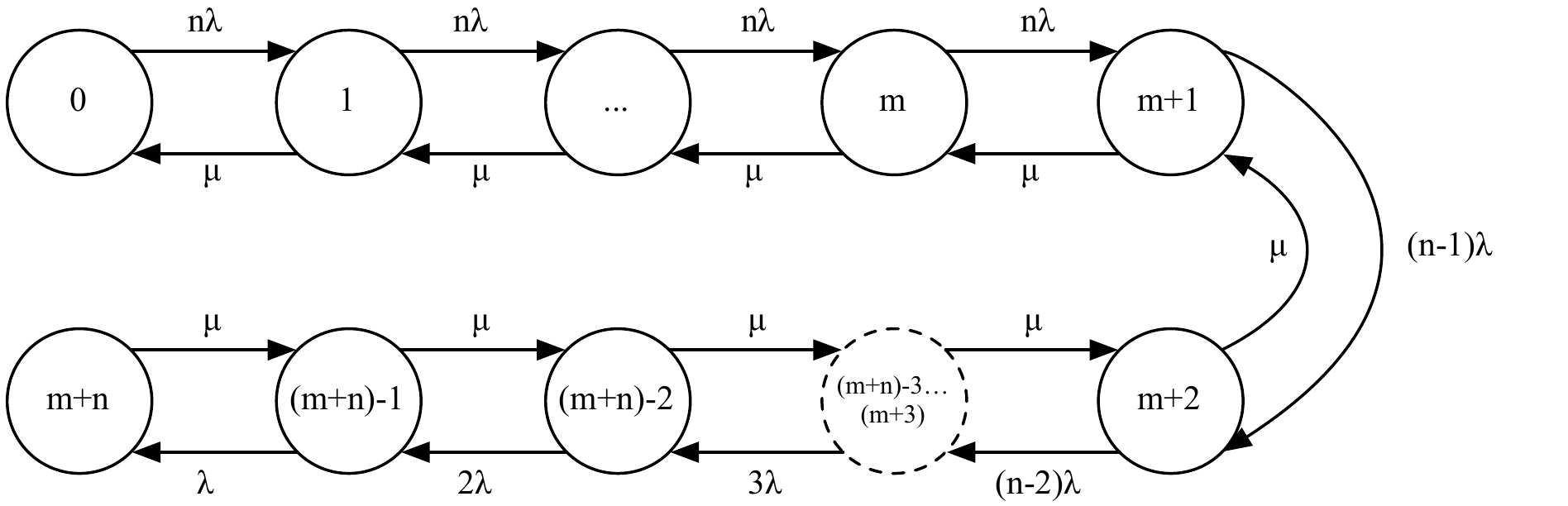}}
\caption{A reference model of a constellation of navigation satellites.}
\label{fig:model2}
\end{figure*}

The reference model of the satellite constellation is depicted in Figure~\ref{fig:model2}. The constellation has $n$ satellites on orbit, and $m$ spare satellites. If the on orbit satellites do not fail, the state of the constellation keeps $n$ satellites available. Once an on orbit satellite fails, one of the spare satellites will replace it immediately to keep $n$ in working condition. If any on orbit satellite fails and there is no spare satellite available to replace it, the number of satellites in the constellation will be reduced to a number smaller than $n$. Thus, spare satellites play a crucial effect on the availability of the satellite constellation.

In the reference model, if the number of satellites in the constellation is $n$ and the number of spare satellites is $m$, where $m\geq 0$ and $n\geq 1$, the launch on schedule (LOS) strategy is to not launch a new satellite. At any time at most one satellite can be repaired. If any on orbit satellite fails, it is immediately replaced by a spare satellite, and repair of the failed satellite commences. If there are no spare satellites, the constellation must operate with fewer than $n$ satellites.

Since the focus of our research is to apply the probabilistic model checking approach and to study its applicability to a satellite constellation, the object of our paper is not limited to any specific navigation satellite system. The system we study here follows a standard configuration for global navigation system. Due to the fact that the current United States' GPS is the most widely used navigation system, parameter values of the constellation also refer to the latest basic parameter settings of such constellation. The parameter values are shown in Table \ref{tab:two}.

\begin{figure}[ht]
\begin{center}
\setlength{\fboxrule}{0.5pt} 
\setlength{\fboxsep}{0.25cm} 
\fbox{\includegraphics[width=12cm]{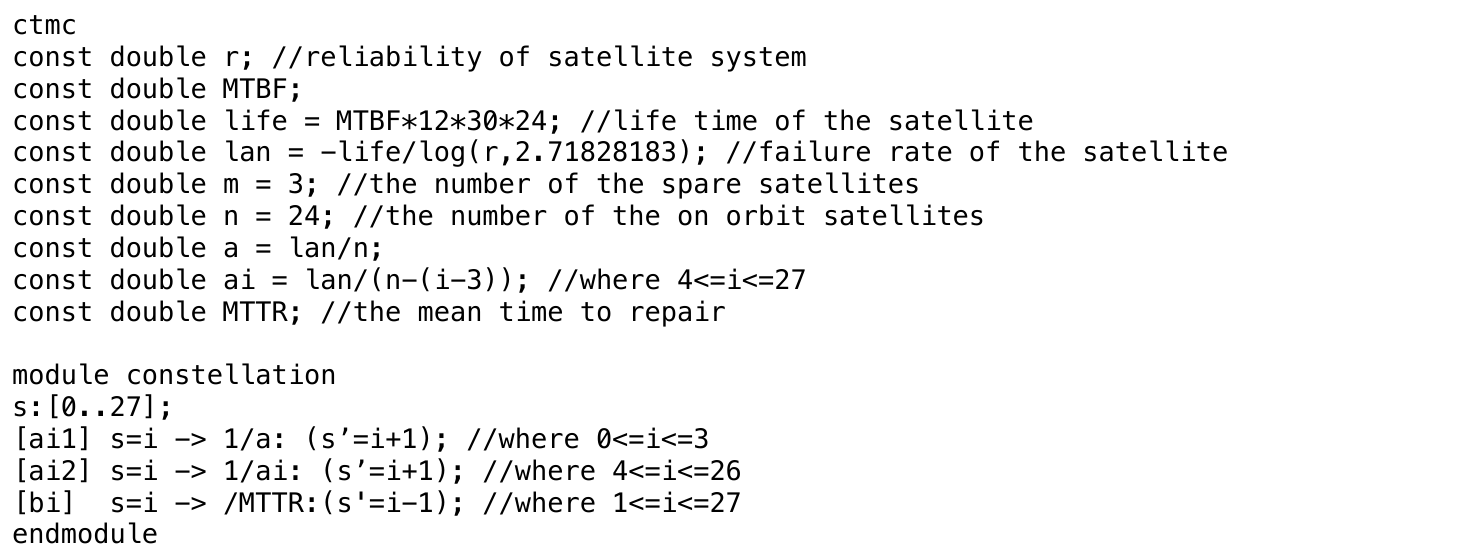}}
\caption{$\mathsf{PRISM}$ module for the satellite constellation.}
\label{fig:code2}
\end{center}
\end{figure}

Our $\mathsf{PRISM}$ specification is given in Figure~\ref{fig:code2}. Assume that the failure and repair rates of a satellite are ${\lambda}$ and ${\mu}$ respectively. When the constellation is operating with $n$ usable satellites, the state transfer rate of the constellation is $n\lambda$. When there are no spare satellites and satellites begin to fail, the transfer rate reduces accordingly to $n\lambda$, where $\lambda$ is the number of functioning satellites.


\section{Quantitative properties and automated analysis}
\label{sect:lspar}

\subsection{Desired properties}

We have identified the need to analyse reliability, availability, and maintainability properties of navigation satellite systems. In the GPS standard proposed in \cite{SIS08}, there are two definitions of availability. The first one is the probability that the slots in the constellation will be occupied by a satellite transmitting a trackable and healthy Standard Positioning Service (SPS) Signal in Space (SIS). The second definition is the percentage of time that the SPS SIS is available to a SPS receiver. According to the same standard, there are two kinds of availability of satellites. The first is the per-slot availability, and the second is the constellation availability, which can be described as follows,

\begin{itemize}
\item Per-slot availability: The time that a slot in the constellation will be occupied by a satellite that is transmitting a trackable and healthy SPS SIS;
\item Constellation availability: the time that a specified number of slots in the constellation are occupied by satellites that are transmitting a trackable and healthy SPS SIS.
\end{itemize}

In our research, we do not consider the environmental effect of the signal for the availability analysis. We only consider fault or failure of satellites. In our context, availability means the ratio of running time for normal satellites to total running time for both normal and failed satellites. The availabilities that we have analysed are: single satellite availability and satellite constellation availability.

The reliability of a satellite depends on planned interruptions, unplanned interruptions, and failure states in the system. The probability of successful launch is the reliability of the satellite, and the maintainability of the satellite is the probability that a satellite can be repaired on orbit. Generally, both reliability and maintainability can be considered as availability properties of the satellite. Reliability must be sufficient to support the mission capability needed in its expected operating environment.

If reliability and maintainability are not adequately designed into satellite systems, there is risk that the design will breach desired availability requirements. Therefore, such system availability baseline is determined by the threshold of design or development costs, which is significantly higher due to resulting corrective action costs. This will cost more than anticipated to use and operate, or will fail to provide the expected availability.

Satellites will deteriorate with time due to failure mechanisms. We assume that time delay is a random variable selected from an exponential distribution, which is an assumption used in $\mathsf{PRISM}$. According to system reliability theory \cite{HR09}, the reliability of a satellite $R(t)$ can be defined as:
\begin{equation}
R(t)=Pr\{T>t\}=e^{-\lambda t},
\end{equation}

\noindent from which we obtain:
\begin{equation}
\lambda(t)=\frac{-lnR(t)}{E(s_{i})}.
\end{equation}

Satellite failures typically occur at some constant failure rate $\lambda$, and failure probability depends on the rate $\lambda$ and the exposure time $t$. According to \cite{Czi13}, typically failure rates are carefully derived from substantiated historical data such as mean time between failure ($MTBF$). We have:
\begin{equation}
\lambda=\frac{-lnR}{T}\ \Longrightarrow\ \lambda=\frac{-lnR}{MTBF},
\label{l4}
\end{equation}
where $t=T=MTBF$, and $MTBF$ is the design parameter or the statistics parameter. Referring to the latest characteristics of satellites used for Global Positioning Systems (GPSs), we assume the $MTBF$ of the satellite to be 15 years. As a result, $R=0.80$ and $MTBF=15\ years$. Further, the mean time to repair ($MTTR$) is $24\ hours$.

\begin{equation}
\mu=\frac{1}{MTTR}.
\label{l5}
\end{equation}

For the evaluation of the availability of the constellation, we focus on long-term failure effects to the constellation. The long term reflect the lifetime of the satellite, and can be described by the $MTBF$ and $MTTR$. The $MTBF$ is used to get the parameter failure rate ${\lambda}$ according to the Equation~\ref{l4}. The $MTTR$ is used to calculate the parameter repair rate ${\mu}$ according to the Equation~\ref{l5}.

\subsection{Formal analysis of a single satellite}

In this section we describe the parameters used in our model and their values. We then use the $\mathsf{PRISM}$ probabilistic model checker to analyse some important properties of the single satellite system. The properties include reliability, maintainability, and availability. The temporal logic CSL is used to analyse the navigation systems because $\mathsf{PRISM}$ supports the use of CSL to verify properties of a CTMC. We then present and analyse our model checking results.

\subsubsection{Reliability properties and results}

Reliability properties of a single satellite that we can analyse using $\mathsf{PRISM}$ include:

\begin{enumerate}
\item when $r=0.80$, the probability that a satellite will need to be replaced by a new one in 15 years:\\
$P_{=?}[F<=T\ s=5];\ T=129600$
\item when $r=0.80$, the probability that a satellite will need to be replaced by a new one due to complete failure in 15 years over time T:\\
$P_{=?}[F<=T\ s=5];\ r=0.80;\ T=0:129600:8640$
\item when $r=0.80$, how many times a satellite will need to be replaced by a new one in 15 years:\\
$R_{=?}[C<=T];\ T= 129600;\ r=0.80$\\
The reward expression in the $\mathsf{PRISM}$ model is the following:\\
$\textbf{rewards}\ ''num\_replace''$\\
$[g]\ true : 1;$\\
$[e]\ true : 1;$\\
$\textbf{endrewards}$
\item how many times a satellite will need to be replaced by a new one over different reliabilities, in 15 years:\\
$R_{=?} [ C<=T ];\ T= 129600;\ r=0.01:0.99:0.05$\\
The reward expression is the same as that for reliability property 3.
\end{enumerate}

In the properties above (and in all other contexts henceforth), 129600 is the lifetime of a satellite in hours (evaluating to approximately $15$ years). Parameter $r$ denotes reliability and proposition $s=5$ asserts that there is a spare satellite on the ground. 
The expression $ r=0.01:0.99:0.05$ indicates that the reliability ranges from $0.01$ to $0.99$ with interval size $0.05$. 

\begin{figure*}[ht]
\begin{minipage}[htbp]{0.5\linewidth}
\centering
\includegraphics[height=4cm]{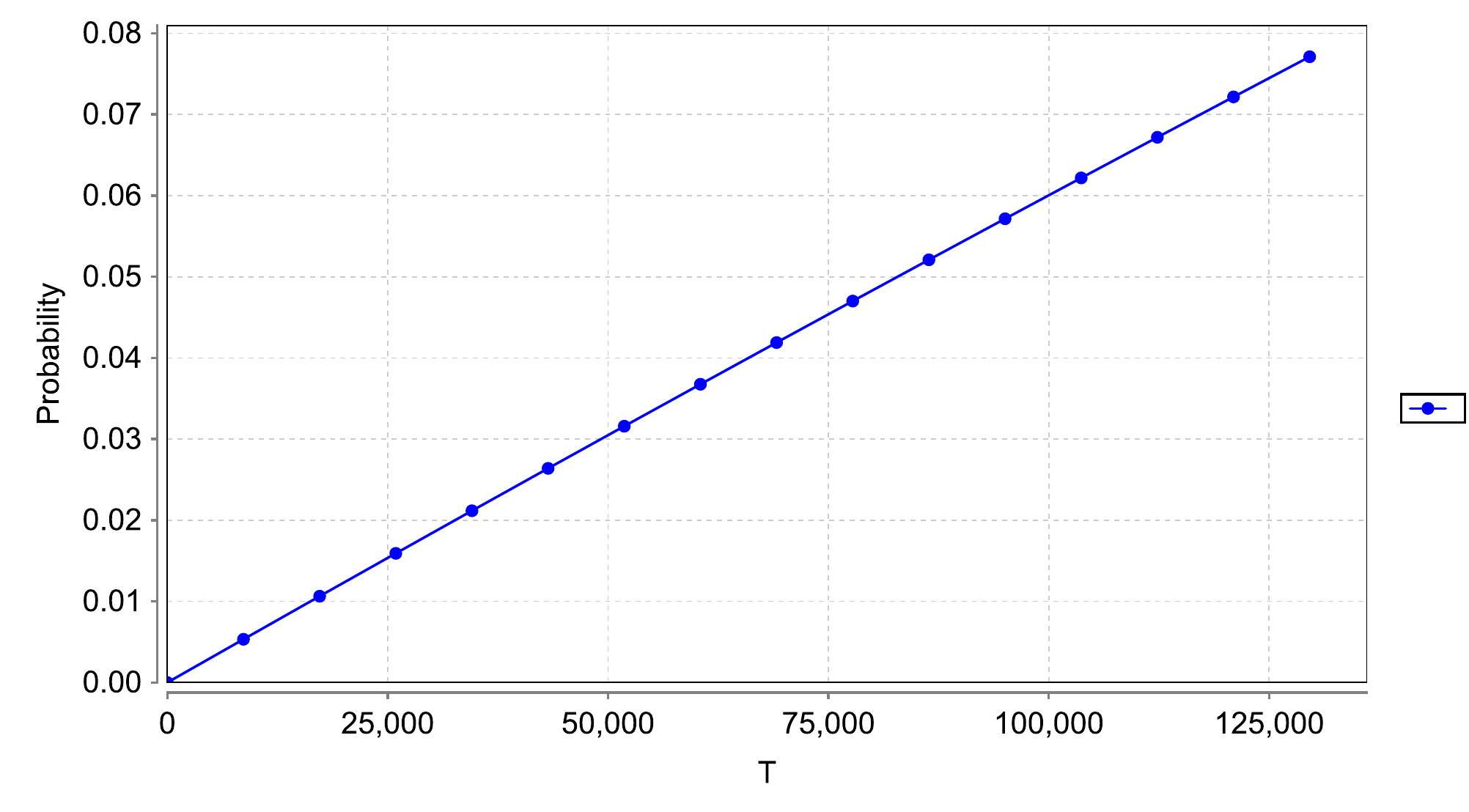}
\footnotesize{\ \ \ (a) Reliability property 2}
\end{minipage}
\begin{minipage}[htbp]{0.5\linewidth}
\centering
\includegraphics[height=4cm]{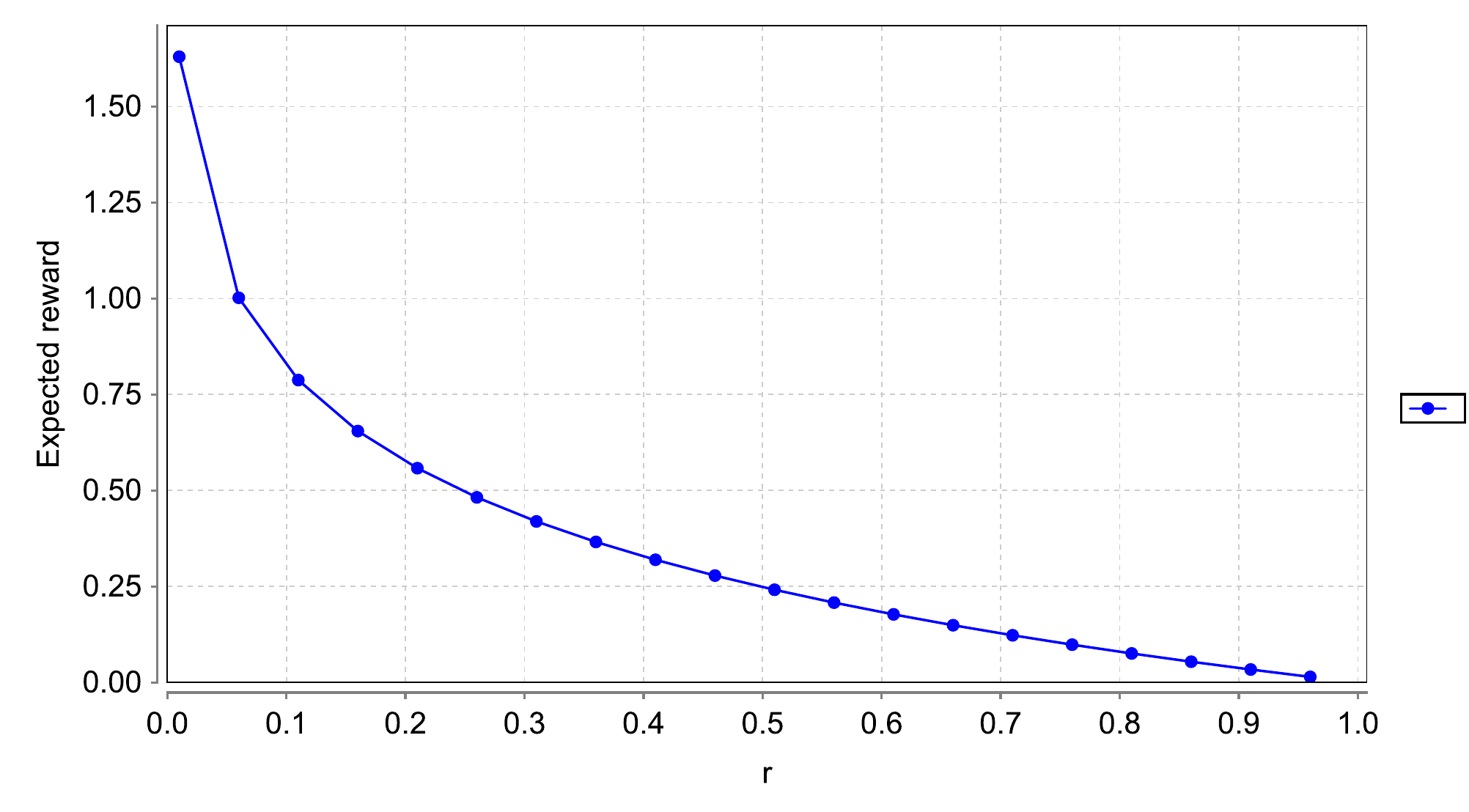}
\footnotesize{\ \ \ (b) Reliability property 4}
\end{minipage}
\caption{Analysis results of reliability properties of a single satellite.}
\label{fig:one}
\end{figure*}

The analysis results of reliability properties which we obtain from $\mathsf{PRISM}$ are as follows. The result of the property 1 is 0.0771; the result of property 2 is shown in Figure~\ref{fig:one}(a); the result of property 3 is 0.08; the result of property 4 is shown in Figure~\ref{fig:one}(b). From Figure~\ref{fig:one}(b), we can see that the number of times the satellite will have a failure and be unable to be repaired in 15 years is 0.08, under the precondition that the reliability is 0.80. If the reliability is set to 0.5, the number of vital failures will be smaller than 0.25 during 15 years. The number of times of unplanned interruptions can be also obtained from the $\mathsf{PRISM}$ by checking the rewards of the unplanned interruption, which is 29.95 times unplanned interruption for the satellite in 15 years.

\subsubsection{Maintainability properties and results}

Maintainability properties of a single satellite that we can analyse using $\mathsf{PRISM}$ include:

\begin{enumerate}
\item when $r=0.80$, the number of times that satellites need to be repaired on orbit in 15 years:\\
$R_{=?} [ C<=T ];\ T=129600;\ r=0.80$\\
The reward expression in $\mathsf{PRISM}$ model is the following:\\
$\textbf{rewards}\ ''num\_repair''$\\
$[d]\ true : 1;$\\
$\textbf{endrewards}$
\item the number of times that the satellite needs maintenance when the reliability is from 0.01 to 0.99 in 15 years:\\
$R_{=?}[C<=T ];\ T=129600;\ r=0.01:0.99:0.01$
\item the number of cases that a satellite needs to be repaired when the $MTBF$ is from 1st year to 15th years:\\
$R_{=?}[C<=T ];\ T=129600;\ r=0.01:0.99:0.01; MTBF=1:129600:8640$\\
The reward expression is the same as that for maintainability property 1.
\item when $r=0.80$, the number of cases that a satellite needs to be repaired on orbit, but not eventually succeed in 15 years:\\
$R_{=?}[C<=T];\ T=129600;\ r=0.80$\\
The reward expression is the same as that for maintainability property 1.
\end{enumerate}

\begin{figure*}[ht]
\begin{minipage}[htbp]{0.5\linewidth}
\centering
\includegraphics[height=4cm]{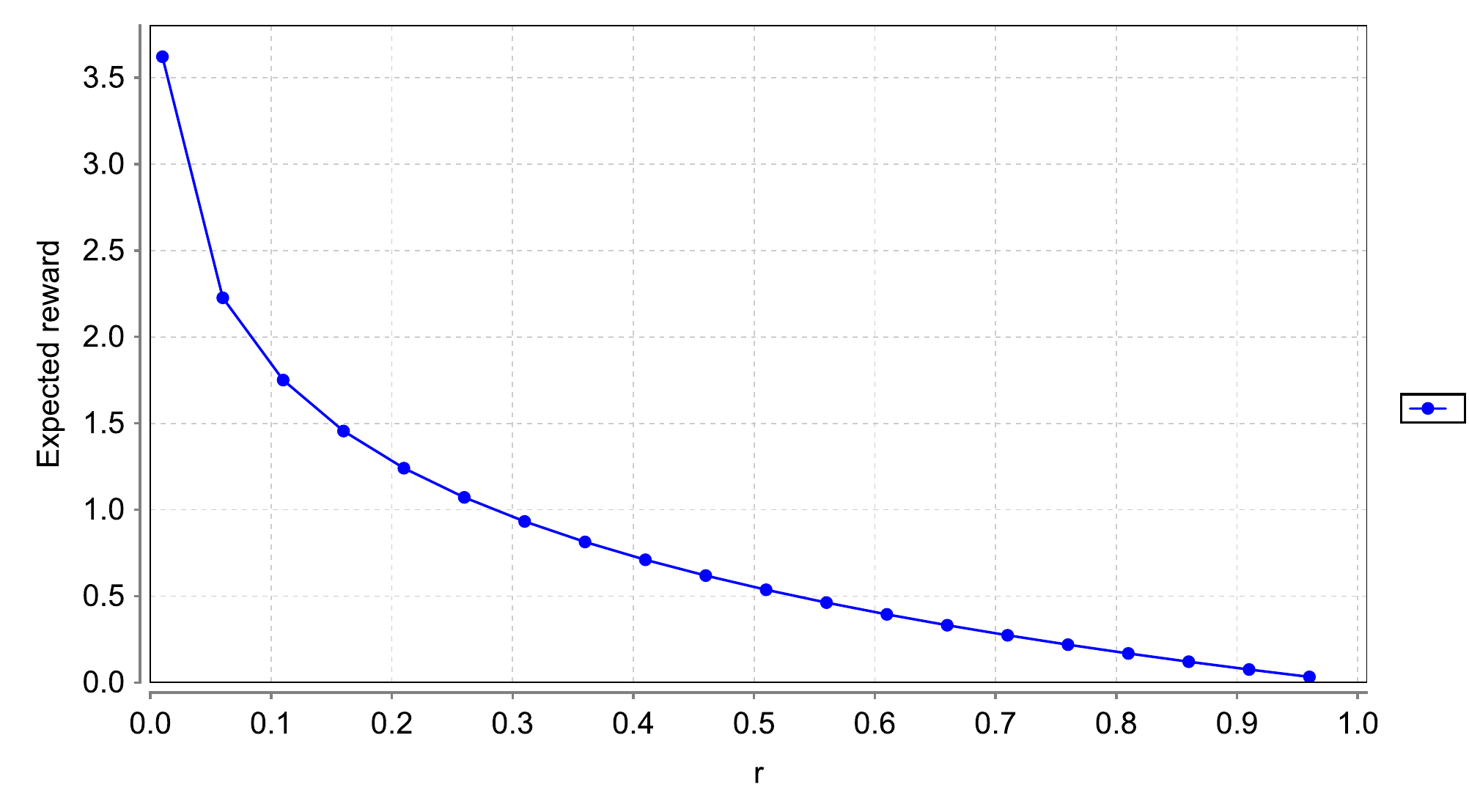}
\footnotesize{\ \ \ \ \ (a) Maintainability property 2}
\end{minipage}
\begin{minipage}[htbp]{0.5\linewidth}
\centering
\includegraphics[height=4.1cm]{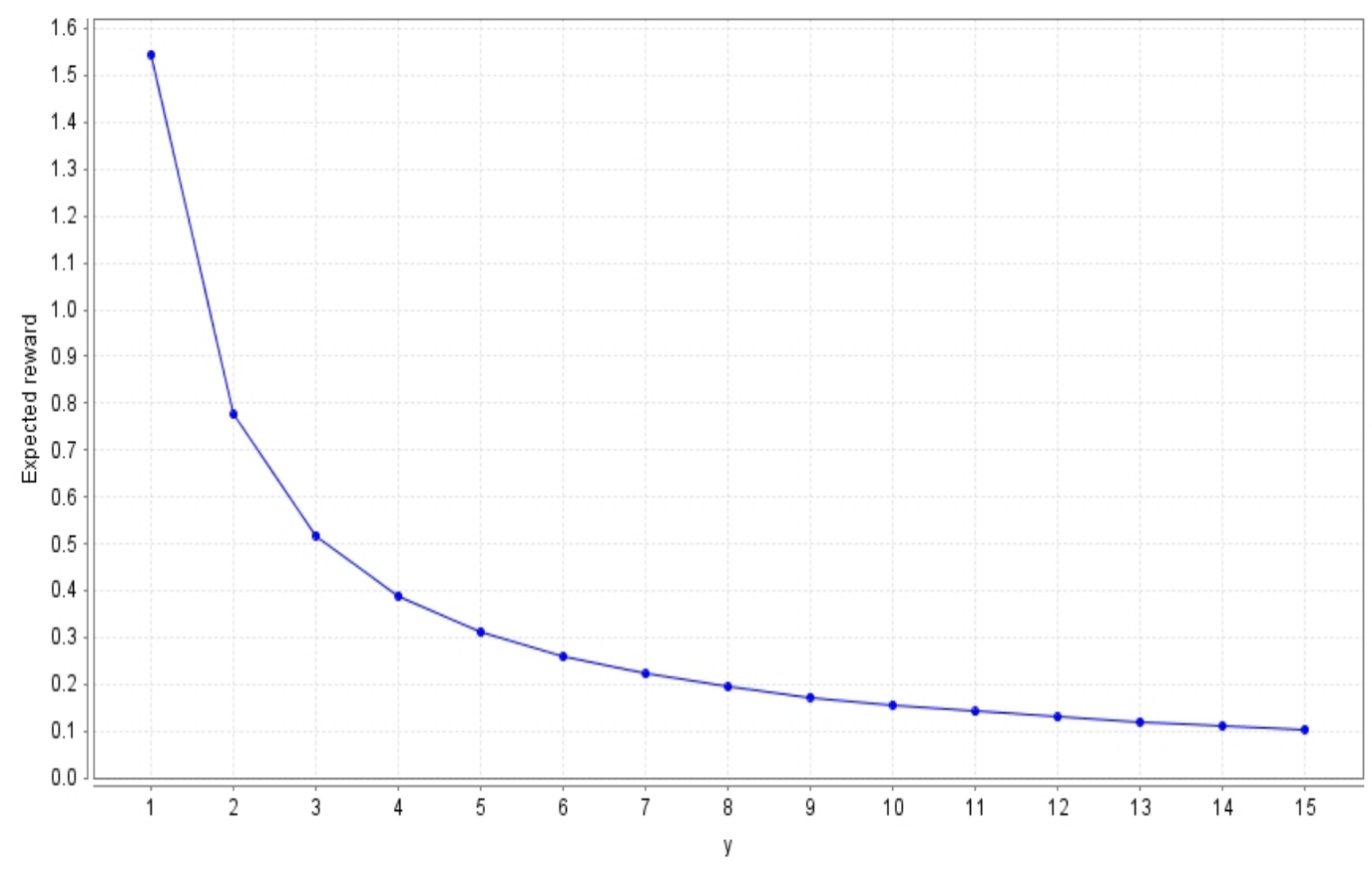}
\footnotesize{\ \ \ \ \ \ \ \ (b) Maintainability property 3}
\end{minipage}
\caption{Analysis results of maintainability properties of a single satellite.}
\label{fig:two}
\end{figure*}

The analysis results of maintainability properties which we obtain from $\mathsf{PRISM}$ are as follows. The result of the property 1 is 0.18; the result of property 2 is shown in Figure~\ref{fig:two}(a); the result of the property 3 is shown in Figure~\ref{fig:two}(b); the result of property 4 is 0.036. The number of times the satellite needs to be repaired on orbit over time is shown in Figure~\ref{fig:two}(a). When the reliability of the satellite is increased to 0.5, the number of times the satellite needs to be repaired will decrease to 0.5. Figure~\ref{fig:two}(b) illustrates that the number of times that the satellite needs to be repaired is below 1 when the $MTBF$ is 2 years. 

\subsubsection{Availability properties and results}

Availability properties of a single satellite that we can analyse using $\mathsf{PRISM}$ includes:

\begin{enumerate}
\item when $r=0.80$, the availability of the satellite in 15 years:\\
$(R_{=?}[C<=T ])/T;\ T=129600;\ r=0.80$\\
The reward expression in $\mathsf{PRISM}$ model is as the following:\\
$\textbf{rewards}\ ''availability''$\\
$s = 0 : 1;$\\
$\textbf{endrewards}$
\item the availability of a satellite over the satellite reliability in 15 years:\\
$R_{=?}[C<=T];\ T=129600;\ r=0.01:0.99:0.01$\\
The reward expression is the same as that for availability property 1.
\item the relationship between satellite availability and its maintenance time taken for planned interruption:\\
$(R_{=?}[C<=T])/T;\ T=129600;\ r=0.80,\ o=1:48:3$\\
The reward expression is the same as that for availability property 1.
\end{enumerate}

\begin{figure*}[ht]
\begin{minipage}[htbp]{0.5\linewidth}
\centering
\includegraphics[height=4.5cm]{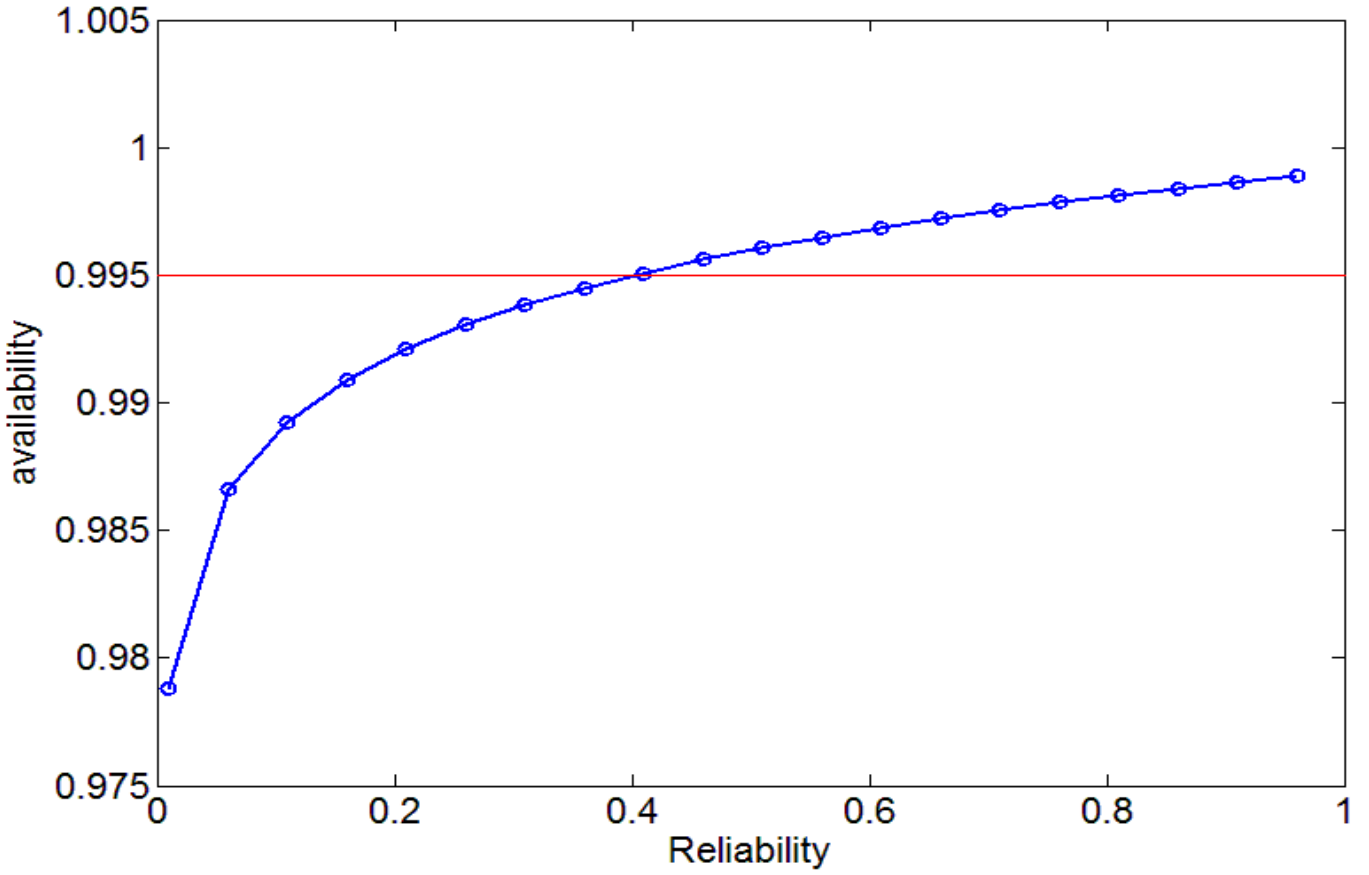}
\footnotesize{\ \ (a) Availability property 2}
\end{minipage}
\begin{minipage}[htbp]{0.5\linewidth}
\centering
\includegraphics[height=4.7cm]{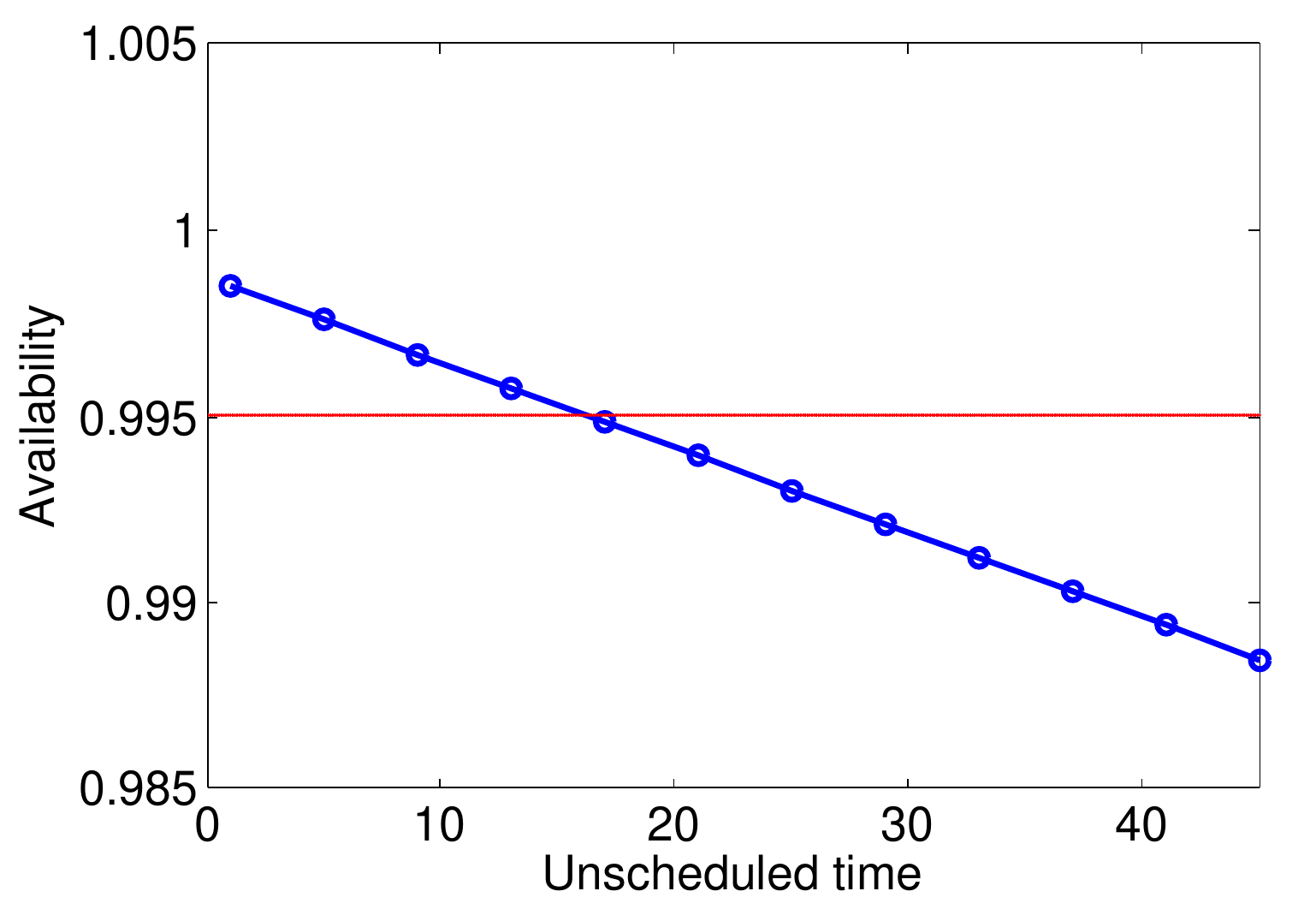}
\footnotesize{\ \ \ \ \ \ (b) Availability property 3}
\end{minipage}
\caption{Analysis results of availability properties of a single satellite.}
\label{fig:three}
\end{figure*}

The analysis results of availability properties which we obtain from $\mathsf{PRISM}$ are as follows. The result of property 1 is 129378 hours; the result of property 2 is shown in Figure~\ref{fig:three}(a); the result of property 3 is shown in Figure~\ref{fig:three}(b). From Figure~\ref{fig:three}(a) we see that if the reliability increases to 0.4, the availability of the satellite reaches 0.995. So if the required probability of the available satellite is 0.995, the reliability must have minimum value 0.4. Figure~\ref{fig:three}(b) indicates that if the required availability is 0.995, the time taken for planned interruption for the satellite will be smaller than 16 hours.

\subsection{Formal analysis of a constellation of navigation satellites}

In this section, we analyse the properties of the satellite system that is made up of a constellation of navigation satellites. Similar to the case of the single satellite, we use $\mathsf{PRISM}$ to check the reliability, maintainability, and availability of the navigation system. We first present properties and their corresponding CSL, and then present and analyse the results of verifying these properties.

\subsubsection{Reliability properties and results}

Reliability properties of the navigation satellite system that we can analyse using $\mathsf{PRISM}$ include:

\begin{enumerate}
\item when the reliability is 0.80, the probability that the number of the useable satellites in the constellation is smaller than 24 in 15 years:\\
$P_{=?}[F<=T\ (s=4)];\ T=129600$
\item when the reliability is 0.80, the probability that the number of the useable satellites in the constellation is smaller than 22 in 15 years:\\
$P_{=?}[F<=T\ (s=6)];\ T=129600$
\item the number of times that all redundant satellites fail in 15 years over the reliability and time:\\
$R_{=?}[C<=T]$\\
The reward expression in $\mathsf{PRISM}$ model is the following:\\
$\textbf{rewards}\ ''num\_fail''$\\
$[a2]\ true:1;$\\
$\textbf{endrewards}$
\end{enumerate}

\begin{figure*}[ht]
\begin{minipage}[htbp]{0.5\linewidth}
\centering
\includegraphics[height=3.5cm]{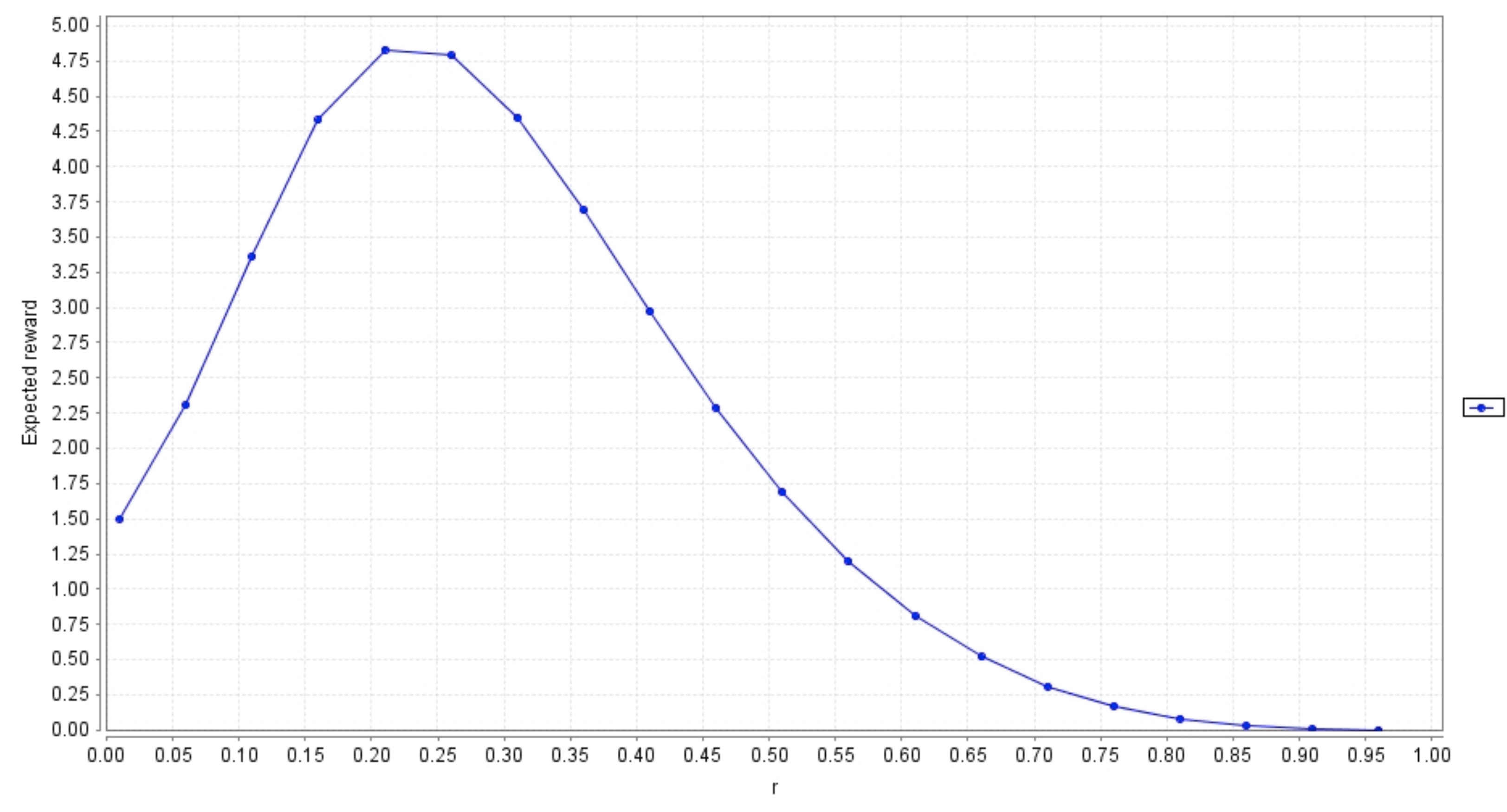}
\footnotesize{(a) Reliability property 3: part 1}
\end{minipage}
\begin{minipage}[htbp]{0.5\linewidth}
\centering
\includegraphics[height=3.6cm]{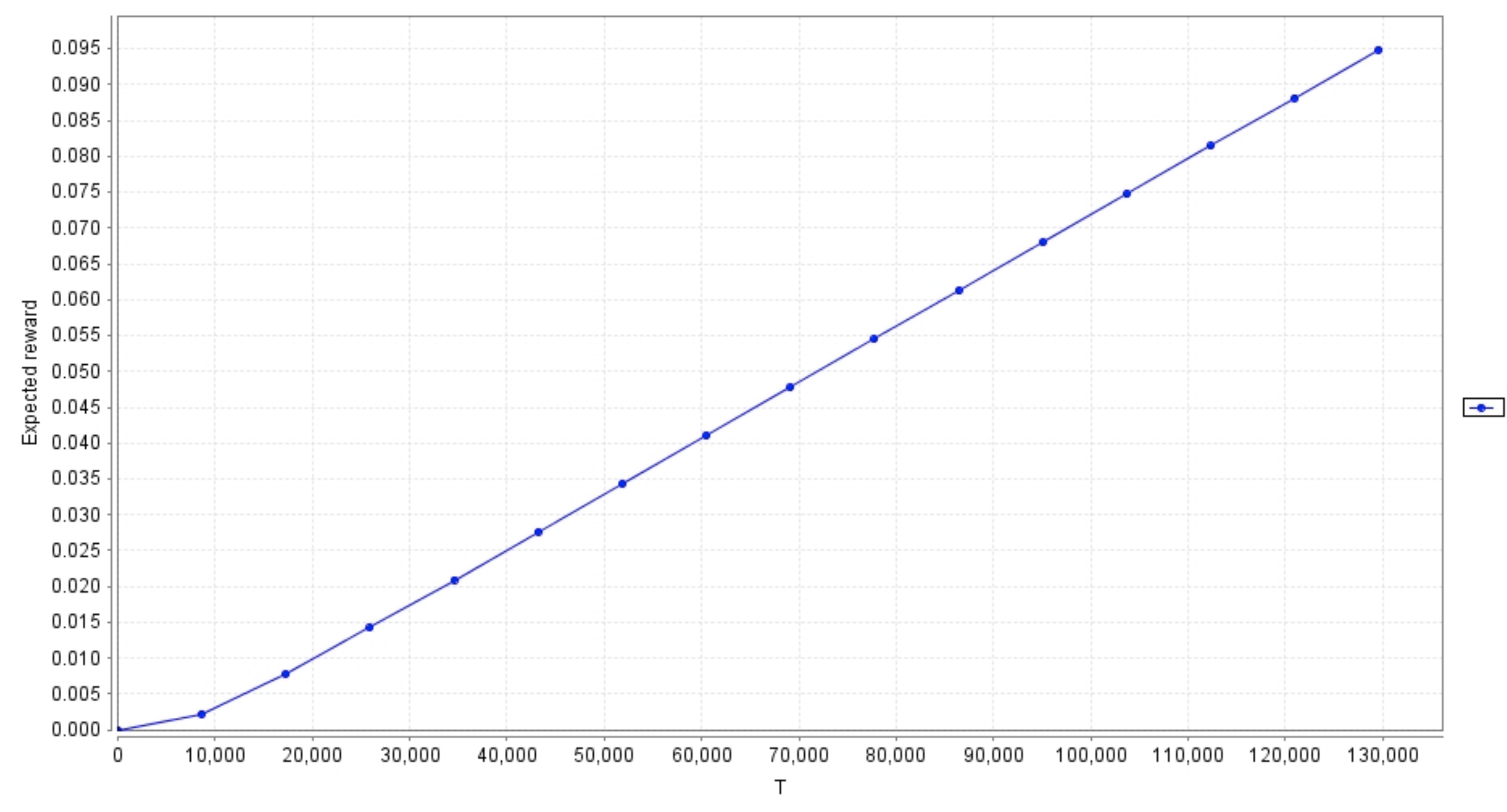}
\footnotesize{(b) Reliability property 3: part 2}
\end{minipage}
\caption{Analysis results of reliability properties of the satellite constellation.}
\label{fig:four}
\end{figure*}

The proposition $s=n$ states that $n$ satellites in the constellation fail.
The analysis results of reliability properties which we obtain from $\mathsf{PRISM}$ are as follows. The result of property 1 is 0.01171; the result of property 2 is 0.0796; the result of property 3 is shown in Figure~\ref{fig:four}.

From Figure~\ref{fig:four}(a), when the reliability is between 0 and 0.25, the number of times that all redundant satellites need to be repaired is proportional to the reliability. As the reliability increases so does the number of required repairs, until the number of repairs reaches $4.76$. However when the reliability is between 0.25 and 1, the number of times that all redundant satellite need to be repaired is inversely proportional to reliability. This is due to the fact that when the reliability decreases to below a specific value, redundant satellites can no longer be repaired. According to Figure~\ref{fig:four}(b), the number of times that all redundant satellites need to be repaired is between 0 and 0.095 in 15 years.

\subsubsection{Maintainability properties and results}

Maintainability properties of the navigation satellite system that we can analyse using $\mathsf{PRISM}$ include:

\begin{enumerate}
\item the average number of times to repair all satellites in the constellation in 15 years:\\
$R_{=?}[C<=T]$\\
The reward expression in $\mathsf{PRISM}$ model is shown as the following:
$\textbf{rewards}\ ''num\_repair''$\\
$[bi]\ true : 1; \\for\ all\ i, 1<=i<=27$\\
$\textbf{endrewards}$
\item The number of times to repair all satellites in the constellation over the reliability in 15 years:\\
$R_{=?}[C<=T];\ r=0.01:0.99:0.05$\\
The reward expression in $\mathsf{PRISM}$ model is the same as that for maintainability property 1.
\item The probability of the case that the number of useable satellites in the constellation is smaller than 22 in 15 years over the number of times for repairing satellites:\\
$P_{=?}[F<=T\ (s=6)];\ T=129600;\ MTTR=0.1:3600:72$
\end{enumerate}

\begin{figure*}[ht]
\begin{minipage}[htbp]{0.5\linewidth}
\centering
\includegraphics[height=3.8cm]{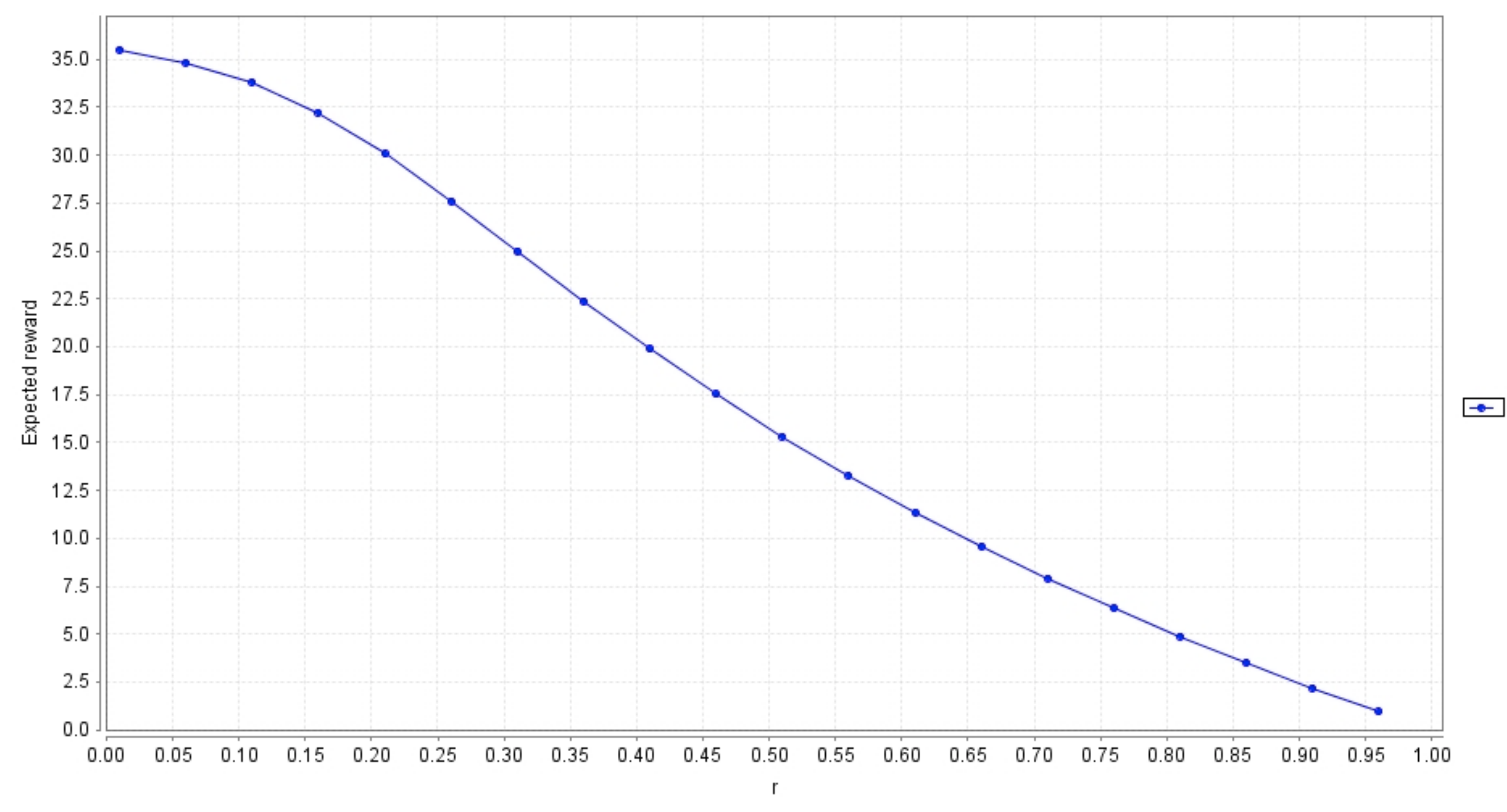}
\footnotesize{\ \ \ (a) Maintainability property 2}
\end{minipage}
\begin{minipage}[htbp]{0.5\linewidth}
\centering
\includegraphics[height=3.8cm]{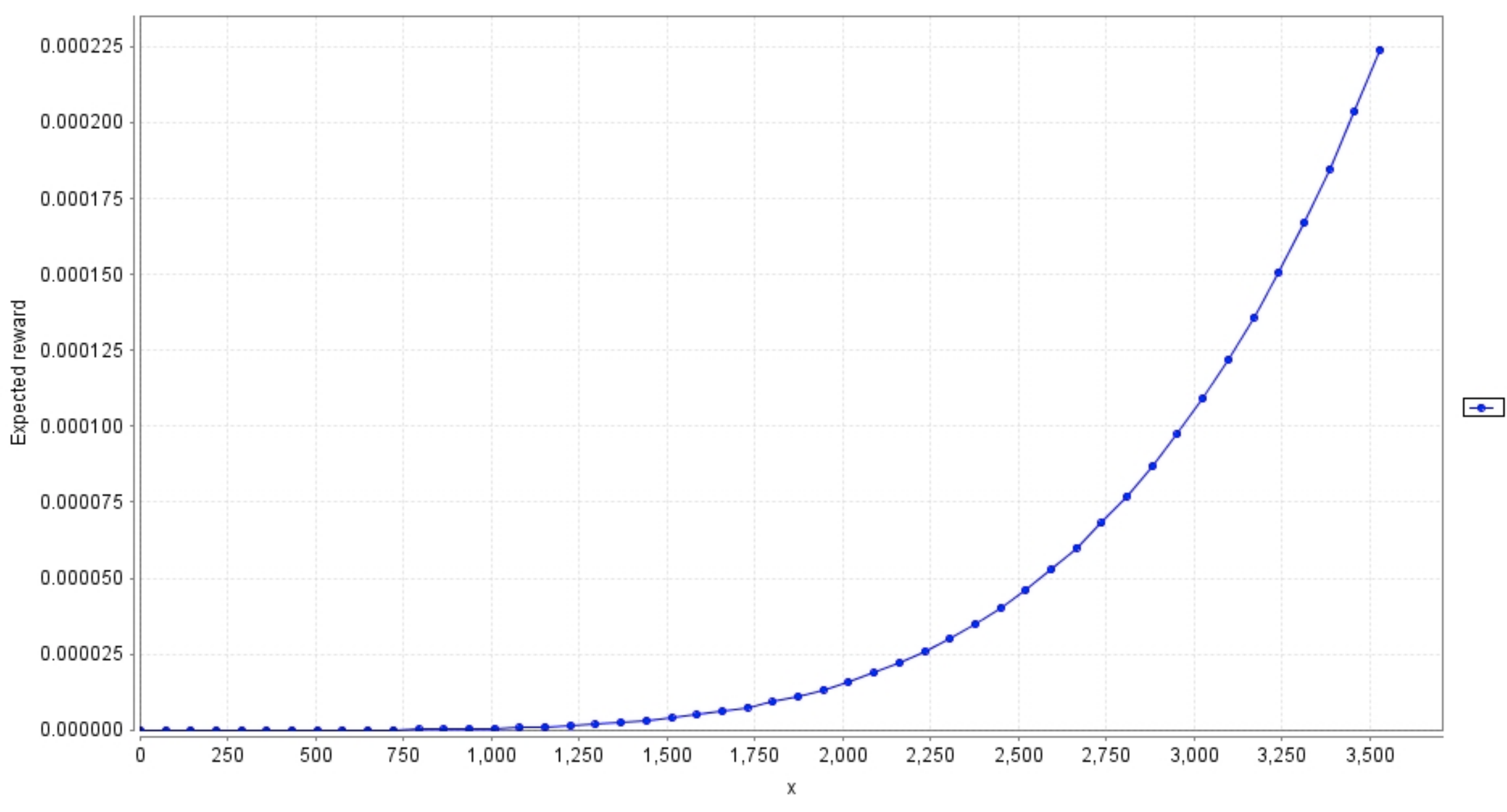}
\footnotesize{\ \ \ (b) Maintainability property 3}
\end{minipage}
\caption{Analysis results of maintainability properties of the satellite constellation.}
\label{fig:five}
\end{figure*}

The analysis results of maintainability properties which we obtain from $\mathsf{PRISM}$ are as follows. The result of property 1 is 5.18; the result of property 2 is shown in Figure~\ref{fig:five}(a) and 
the result of the property 3 is shown in Figure~\ref{fig:five}(b).

From Figure~\ref{fig:five}(a) we see that as reliability increases, the number of times that all satellites in the constellation need to be repaired over 15 years decreases from 35 to 2.5 when the reliability reaches 0.90. As depicted in Figure~\ref{fig:five}(b), the probability that the constellation consists of $n$ satellites with $n$ is smaller than 22 in 15 years is 0.0225.

\subsubsection{Availability properties and results}

Availability properties of the navigation satellite system that we can analyse using $\mathsf{PRISM}$ include:

\begin{enumerate}
\item the period of time that the constellation consists of 24 satellites in 15 years:\\
$R_{=?}[C<=T];$,\\
and reward expression in the $\mathsf{PRISM}$ model is shown as below:\\
$\textbf{rewards}\ ''reward''$\\
$s=i:1;$ where $0<=i<=3$\\
$\textbf{endrewards}$
\item the availability of the constellation consists of 24 satellites in 15 years:\\
$(R_{=?}[C<=T])/T;$,\\
and reward expression is the same as the availability property 1;
\item the availability of the constellation consists of 24 satellites in 15 years over the reliability:\\
$(R_{=?}[C<=T])/T;\ r=0.01:0.99:0.05$,\\
and reward expression is the same as the availability property 1;
\item the availability of the constellation consists of 24 satellites in 15 years over the repair time:\\
$(R_{=?}[C<=T])/T;\ MTTR=0.1:3600:72$\\
and reward expression is the same as the availability property 1.
\end{enumerate}

\begin{figure*}[ht]
\begin{minipage}[htbp]{0.5\linewidth}
\centering
\includegraphics[height=4.7cm]{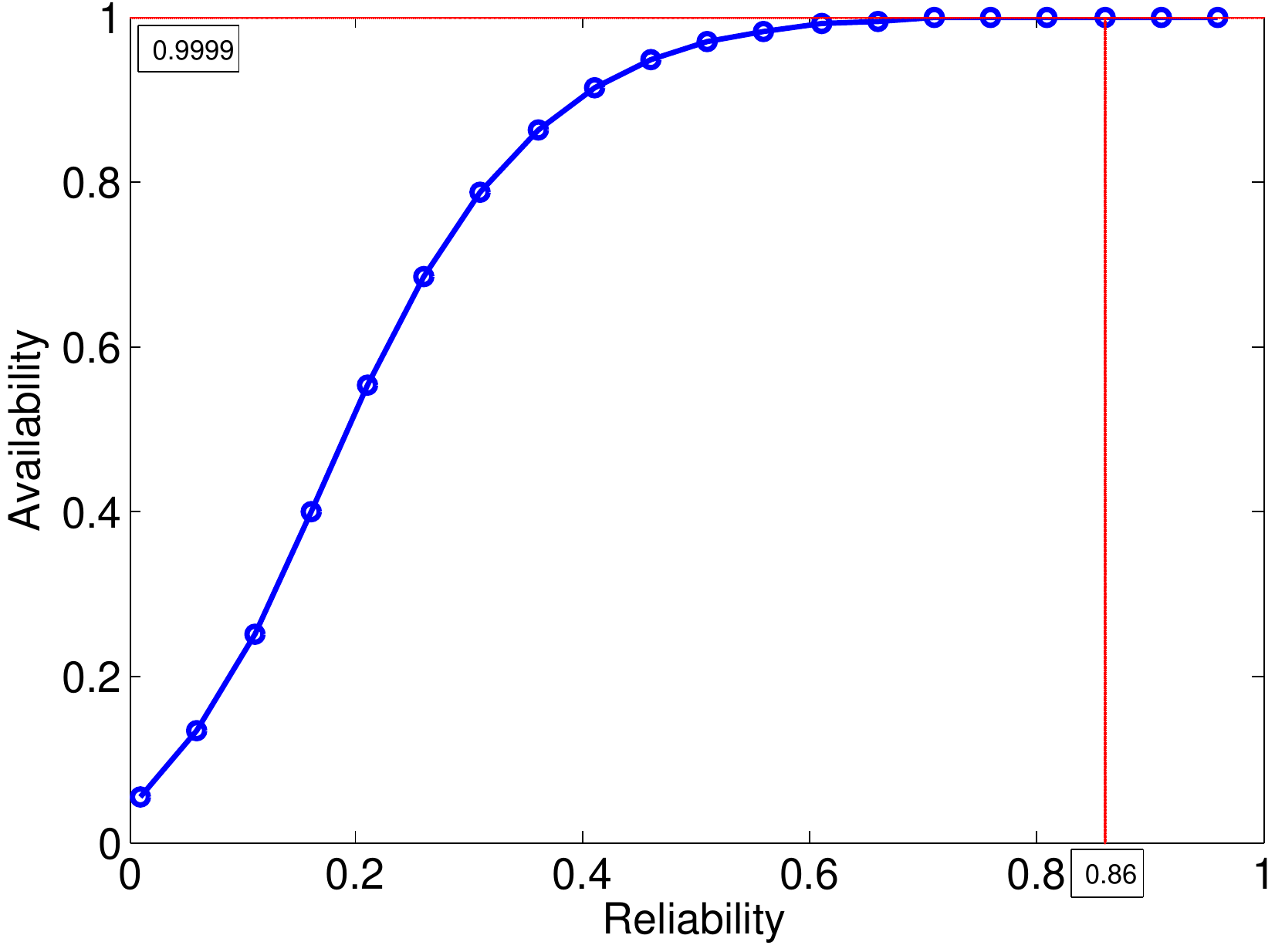}
\footnotesize{\ \ \ \ \ \ (a) Availability property 3}
\end{minipage}
\begin{minipage}[htbp]{0.5\linewidth}
\centering
\includegraphics[height=4.9cm]{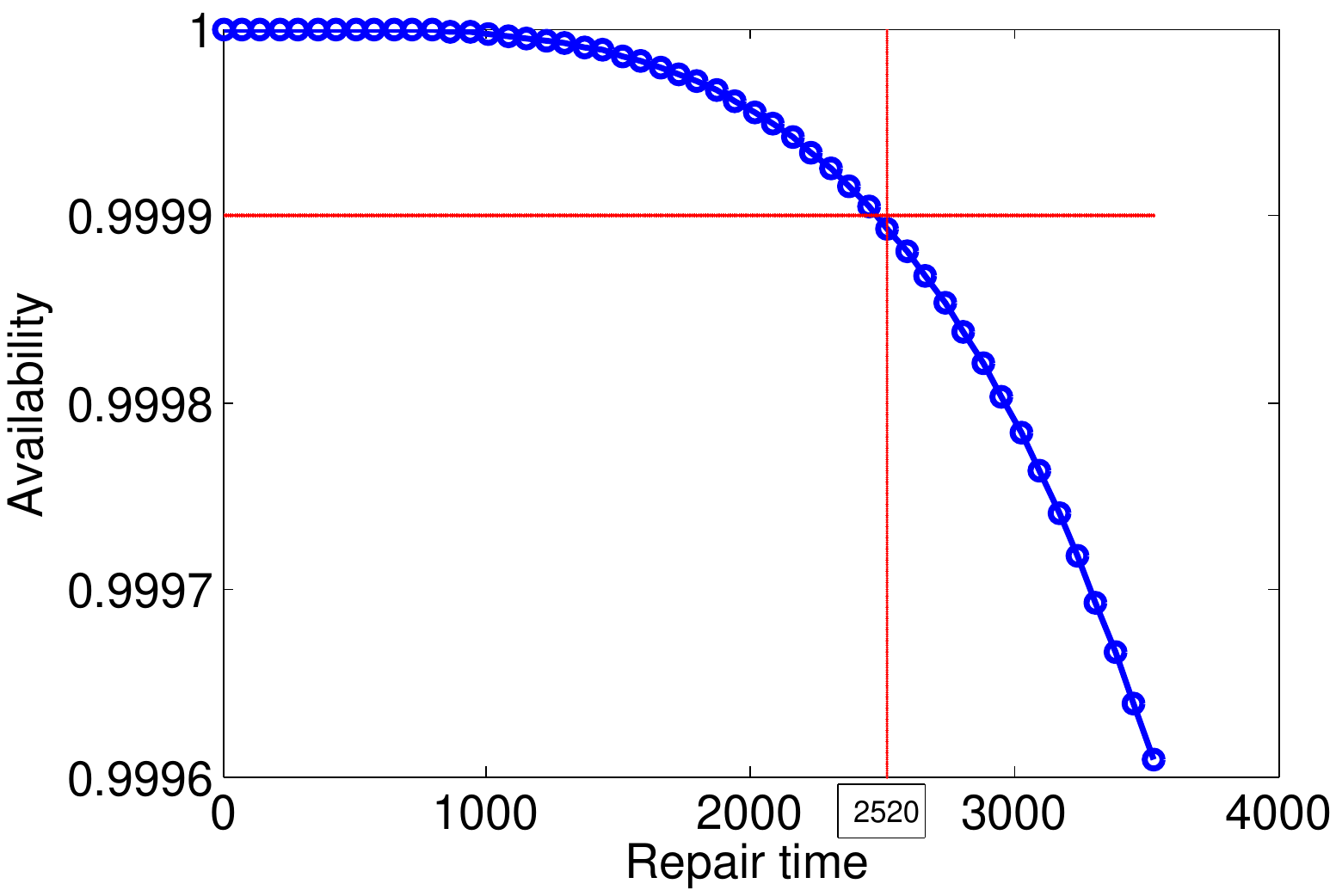}
\footnotesize{(b) Availability property 4}
\end{minipage}
\caption{Analysis results of availability properties of the satellite constellation.}
\label{fig:six}
\end{figure*}

The analysis results of availability properties which we obtain from $\mathsf{PRISM}$ are as follows. The result of property 1 is 129545 hours; the result of property 2 is 0.99958; the results of properties 3 and 4 are shown in Figures~\ref{fig:six}(a) and \ref{fig:six}(b) respectively.

The availability of the satellite constellation as the reliability and time taken to repair satellites increases is shown in Figures~\ref{fig:six}(a) and \ref{fig:six}(b) respectively. According to Figure~\ref{fig:six}(a), if the availability of the constellation is 0.9999 and the time taken to repair a satellite is 5 months, the reliability is at least 0.86. When the reliability is 0.80 , for the same availability requirement of the constellation, when the satellite has a fault or fails, the time taken to repair a satellite is at most 2520 hours (3.5 months).

\subsection{Discussion of results}

Since parameter settings of our formal models are based on GPS Block III which is the newest generation of GPS systems, our analysis results can be compared to existing GPS statistical analysis. According to a report of Lockheed Martin \cite{Jac13}, a leading global security and aerospace company, the availability of the GPS Block III is given as 99.9\%. The availability we evaluate in this paper is close to the actual data. According to a further Lockheed Martin report \cite{Sha11}, the constellation availability of the GPS Block III is given as 99.88\%.

In this paper, the availability we evaluate for two scenarios is in each case close to the actual data. This has proved to be both useful and efficient to use probabilistic model checking approach for the modelling and analysis of a singe satellite and a constellation of navigation satellites. To the best of our knowledge, we are the first to use the formal technique of probabilistic model checking to perform RAM analysis of satellite systems. These results indicates that our approach can also be applied to a wider range of quantitative properties of formal models taken from many application domains for satellite systems.

\subsection{Benefits of the approach}

To address the performance of satellite systems, it is essential to accurately quantify aspects such as reliability, availability and maintainability. There are two common techniques can be used for evaluating these features. One is the reliability block, and the other is the fault tree. However, neither technique is suitable to evaluate probabilistic properties, due to the fact that they are static techniques. In a fault tree or reliability block formalism, it is necessary to assume the probabilities of each fault or failure are independent, while this is not the case in reality.


Other benefits of applying probabilistic model checking with $\mathsf{PRISM}$ for the specification and analysis of satellite systems is that the results can be plotted as graphs that can be inspected for trends and anomalies. Furthermore, we are able to compute exact quantities, rather than approximations based on a large number of simulations, thus enabling us to obtain complete and exhaustive conclusions for all possible parameter values. In addition, $\mathsf{PRISM}$ enables automated analysis. This helps manual analysis with automatic analysis support, thus making development more efficient and minimising human errors during the design phase.

There are also some disadvantages to using Markov models, not least that their specification, and the specification of useful properties, requires a high degree of mathematical skill. Markov models may be large and cumbersome in some cases, and the specification can be error-prone. In addition, as a system increases in complexity, so does the size of the state-space associated with a corresponding model. This results in a longer (possibly intractable) search.


\section{Related work}
\label{sect:relatedWork}

There have been a number of notable attempts to use formal methods to address the problems of design exploration for a satellite system. The theorem prover PVS \cite{ORS92} was used to verify desired properties in system models of Ariane 5 where the cost of failure is high. The PICGAL project \cite{DLV97} analysed ground-based software for launch vehicles similar to Ariane 5. In a NASA report \cite{Rus97}, formal methods and their application to critical systems are explained to stakeholders from the aerospace domain. In \cite{CD04} Markov models are used to evaluate the cost of availability of coverage of satellite constellation. The potential role of formal methods in the analysis of software failures in space missions is discussed in \cite{Joh05}, .

Similarly, \cite{BDG+06} explores how verification techniques, such as static analysis, model checking, and compositional verification, can be used to gain trust in space-based systems. Model checking has proved to be a suitable formal technique for exposing errors in satellites, mainly due to classical concurrency errors. Unforeseen interleaving between processes may cause undesired events to occur. In \cite{HLP01}, the SPIN model checker \cite{Hol04} was used to formally analyse a multi-threaded plan execution module, which is a component of NASA's artificial intelligence-based spacecraft control system as a part of the Deep Space 1 mission. Five previously undiscovered errors were identified in the spacecraft controller, in one case representing a major design flaw.

The model checker Mur$\psi$ \cite{DDH+92} has been used in \cite{She01} to model the Entry, Descent and Landing phase of the Mars Polar Lander. It was then used to search for sequences of states that led to the violation of a Mur$\psi$ invariant. This stated that the thrust of the pulse-width modulation, which controls the thrust of the descent engines, should always be above a certain altitude. In \cite{GDH13} the model checker NuSMV \cite{CCG+02} is used to model and verify the implementation of a mission and safety critical embedded satellite software control system. The control system is responsible for maintaining the altitude of the satellite and for performing fault detection, isolation, and recovery decisions, at a detailed level.

Furthermore, model checking is used in \cite{CMM+11} to simulate satellite operational procedures, and it exploits a simulator of a satellite as a black box in order to formally verify operational procedures. In \cite{CMM+11, GCM+12}, exhaustive search of all possible simulation scenarios is performed, using the simulator as a model. Thus the verification is automated and complete. Moreover, the approach of system level formal verification to exploit a simulator in order to carry out formal verification has been further developed in \cite{MMM+13,MMM+14} and applied to biological contexts. Finally, all these approaches use the explicit model checker CMurphi \cite{PIM+04}.

Our preliminary research into the verification of satellite systems, in which we restrict our analysis to a single satellite, is presented in \cite{PLM+13}. In work \cite{EKY+12} similar to ours, formal techniques have been used on a regular design of a modern satellite. In that work the COMPASS automated tool is used to carry out their analysis. COMPASS \cite{BCK+11} supports model checking techniques for verifying correctness, using fault trees for safety analysis. The major difference between this work and ours is that we perform formal analysis of quantitative properties such as reliability, availability, and maintainability of both a single satellite and a constellation of satellites. Whereas, \cite{EKY+12} mainly verifies qualitative properties (such as correctness, safety, and dependability) of a single satellite.


\section{Conclusions and future work}
\label{sect:conclusion}

Reliability, availability and maintainability (RAM) analysis of systems has been indispensable in the design phase of satellites in order to achieve minimum failures or to increase mean time between failures ($MTBF$) and thus to plan maintainability strategies, optimise reliability and maximise availability. Traditional approaches are not suitable for performing RAM analysis of navigation satellite systems. We present formal models of both a single satellite system and a constellation of navigation satellites and logical specification of reliability, availability and maintainability properties. We have analysed a set of properties using the automatic probabilistic model checker $\mathsf{PRISM}$.

There are many technical and theoretical challenges that remain to be addressed. In particular, satellite failure often forms part of more complex problems that show through different aspects of the engineering of space based systems. The technical challenges also include basic issues with the representation of safety and space mission critical characteristics of satellite telecommunications due to a group of satellites working together given the limitations of classical modelling approach.

In order to fully explore satellite behaviour, it will be necessary to exploit further formal techniques. For instance, if we want to model the mobility of connection between satellites it may be necessary to express behaviour via an extension to the $\pi$-calculus, and model check using $\mathsf{PRISM}$ (a technique identified in \cite{NPP+09}). This kind of issue must be addressed in order to identify the causes of satellite system failure and to support the development of satellite systems.

As $\mathsf{PRISM}$ assumes events to occur according to an exponential distribution, we are limited to making the same assumption about the events in our systems. In fact, many types of satellite failure follow a different distribution. In particular, a number of failures of satellites have a Weibull distribution \cite{Bir10}, which follows the conventional three-component bathtub curve which models a burn-in and wear-out phase for failure prediction. For future work, we will look at how to represent arbitrary distributions in probabilistic models, and to what extent such kind of distributions are able to be supported by the probabilistic model checking approach.


\section*{Acknowledgement}

The authors would like to thank the editors for their help and referees for their valuable comments. This research was partially supported by the EC under the EATS project: European Train Control System Advanced Testing and Smart Train Positioning System (FP7-TRANSPORT-314219). The author Yu Lu was funded by the Scottish Informatics and Computer Science Alliance (SICSA) under the research theme of Modelling and Abstraction.

\bibliographystyle{qrei}
\bibliography{qrei}

\clearpage

\section*{Tables}

\begin{table*}[ht]
\centering
\begin{tabular}[l]{@{}ccccccccccc}
\hline
  $r$ & $MTBF$ & $MTTR$ & $t_{u}$ & $t_{p}$ & $p_{b}$ & $t_{r}$ & $t_{d}$ & $t_{e}$ & $p_{y}$ & $t_{k}$ \\
 & years & hours & hours & hours & & hours & hours & hours & & hours\\
\hline
  0.80 & 15 & 24 & 4320 & 4320 & 0.80 & 24 & 1440 & 4320 & 0.90 & 24\\
\hline
\end{tabular}
\caption{Parameters used in the model for the single satellite system .}
\label{tab:one}
\end{table*}

\begin{table*}[ht]
\centering
\begin{tabular}[l]{@{}ccccc}
\hline
  $r$ & $MTBF (T)$ & $MTTR$ & $n$ & $m$ \\
\hline
   0.80 & 15 years & 5 months & 24 & 3 \\
\hline
\end{tabular}
\caption{Parameters for the navigation satellite systems.}
\label{tab:two}
\end{table*}

\clearpage

\section*{Figures}

\end{document}